\pdfoutput=1
\documentclass[useAMS, usenatbib]{mn2e}
\usepackage[]{natbib}
\usepackage[T1]{fontenc}
\usepackage{aecompl}
\usepackage{rotating}
\usepackage{stmaryrd}
\usepackage{aas_macros}
\usepackage{times}
\usepackage{graphicx,subfig}
\usepackage{amsmath}
\usepackage{amssymb}
\usepackage{amsfonts}
\usepackage{placeins}

\def\gsim { \lower .75ex \hbox{$\sim$} \llap{\raise .27ex \hbox{$>$}}}
\def\lsim { \lower .75ex \hbox{$\sim$} \llap{\raise .27ex \hbox{$<$}}}
\newcommand\ion[2]{\ensuremath{\mathrm{#1\,\scriptstyle #2}}}

\begin{document}
\newcounter{MYtempeqncnt}
\graphicspath{{fig/}}

\title{The broadening of Lyman-$\alpha$ forest absorption lines}

\author[A. Garzilli, T. Theuns and J. Schaye]  {Antonella Garzilli$^{1}$\thanks{E-mail:
    garzilli@lorentz.leidenuniv.nl}, 
  Tom Theuns$^{2}$ and
  Joop Schaye$^{3}$\\
  $^1$ Lorentz Institute, Leiden University, Niels Bohrweg 2,
  Leiden, NL-2333 CA, The Netherlands\\
  $^2$ Institute for Computational Cosmology, Department of Physics, Durham University, DH1 3LE Durham, UK \\
  $^3$ Leiden Observatory, Leiden University, P.O. Box 9513, 2300 RA Leiden, The Netherlands
} \date{Accepted 2015 February 20. Received 2015 February 19; in
  original form 2014 December 28}

\maketitle

\begin{abstract}
We provide an analytical description of the line broadening of \ion{H}{I} absorbers
in the Lyman-$\alpha$ forest resulting from Doppler broadening and Jeans smoothing.
We demonstrate that our relation captures the dependence of the line-width on column density for
narrow lines in $z\sim 3$ mock spectra remarkably well. Broad lines at
a given column density arise when the underlying density structure is
more complex, and such clustering is not captured by our model.  Our
understanding of the line broadening opens the way to a new method to
characterise the thermal state of the intergalactic medium and to determine the sizes of the absorbing structures.
\end{abstract}


\section{Introduction}
The intergalactic medium (IGM) is detected as intervening absorption in the spectra of background sources
\citep{gunn1965}, in the form of a \lq forest\rq\ of \ion{H}{I} (\cite{lynds1971, weymann1981}, see {\em e.g.} \cite{rauch1998} for a review) and \ion{He}{II} \citep{jakobsen1994,davidsen1996} absorption lines. This gas is very highly ionised, $x=n_{\rm HI}/n_{\rm H}\sim 10^{-4}$, by a pervasive background of ionising radiation produced by galaxies and quasars 
\citep[\protect{\em e.g.}][]{haardt1996}, with hints of an increasingly neutral fraction, $x\sim 0.1$, above redshift $z\sim 7$ \citep{mortlock2011}. At all times, it contains the majority of baryons \citep{fukugita1998}.

The growth of structure in a cold dark matter (CDM) Universe naturally
gives rise to a \lq cosmic web\rq\ of voids, sheets and filaments,
with absorbers arising when the sight-line intersects higher density
regions, as demonstrated in analytic models \citep{bi1997,schaye2001}
and simulations \citep{cen1994,weinberg1996,theuns1998}. The
temperature of this gas is set by the balance between photo-heating and adiabatic expansion and compression, resulting in a well-defined
temperature-density relation,
$T=T_0(\rho/\langle\rho\rangle)^{\gamma-1}$ , where $\gamma\approx 1$
immediately after reionisation, and $\gamma\rightarrow 1+1/1.7$
asymptotically long after reionisation \citep{huignedin1997,theuns1998}. Patchy reionisaton of either \ion{H}{I} or \ion{He}{II} may
lead to patchy photo-heating \citep{abel1999,mcquinn2009} and hence
spatial fluctuations in the temperature, but these have not (yet) been
detected \citep{theuns2002,mcquinn2011}.

Feedback from star formation is an essential ingredient in models of
galaxy formation and might impact the IGM \citep{theuns2001}. For
example, elements synthesised in stars are detected in the IGM
\citep{cowie1995}, even at low densities \citep{schaye2003}. If, as is
likely, this enrichment is due to galactic winds, then the IGM might
not be ideal for making {\em cosmological} inferences. Fortunately, it
appears that the effects of such winds are relatively small
\citep{theuns2002c, mcdonald2005, viel2013}. However, current
simulations still struggle somewhat to reproduce the detected
enrichment patterns as a function of density \citep{aguirre2005,wiersma2010,wiersma2011} and hence may underestimate the effects of galaxy formation on the IGM.

Accurate measurements of the evolution of $T_0$ and $\gamma$ are
interesting, since they could constrain the epoch of \ion{H}{I} and
\ion{He}{II} reionisation \citep{theuns2002b}, but also because the
formation of small galaxies is quenched due to a reduction of atomic
cooling \citep{efstathiou1992} following reionisation, and because
photo-heating evaporates the gas out of shallow dark matter potentials
\citep{okamoto2008}. This is a crucial ingredient in models of galaxy
formation, in particular for dwarf galaxies and satellites
\citep{benson2002, sawala2014}. The thermal state of the IGM is also
an important \lq nuisance\rq\ factor when using the Lyman-$\alpha$
forest to study the nature of the dark matter \citep{boyarsky2009}, or
to determine the amplitude of the UV-background \citep{bolton2005}.

\cite{schaye1999} described a method for inferring $(T_0,\gamma)$ from the line width versus column density ($b-N_{\rm HI}$) scatter plot of Lyman-$\alpha$ lines, when the Lyman-$\alpha$ forest is fitted as a sum of Voigt profiles for example using {\sc vpfit}\footnote{{\sc vpfit} is developed by R. Carswell and J. Webb, http://www.ast.cam.ac.uk/\~rfc/vpfit.html.}. This improved and extended the work of \cite{theuns1998} who had noted that the small $b$ cut-off of the $b$-parameter distribution was sensitive to temperature. \cite{schaye2000}, and \cite{ricotti2000, bryan2000} and \cite{mcdonald2001}
used this to measure the $T-\rho$ relation at redshifts $z\sim 2-4$.

Since then other measurements have been performed over a larger redshift
range, some also based on Voigt profile fitting \citep{rudie2012},
or based on small-scale power spectrum or wavelet analysis \citep{theuns2000a,
  mcdonald2000, zaldarriaga2001, theuns2002b, viel2009,
  lidz2010,garzilli2012}, transmission PDF statistics
\citep{bolton2008, viel2009, calura2012, garzilli2012} and more
recently on curvature statistics \citep{becker2011}. Current
measurements still have large error bars for $T_0$ and constraints are
even poorer for $\gamma$. \cite{bolton2008} and \cite{viel2009} even
argue that $\gamma<1$, which is difficult to understand from a
theoretical point of view \citep{mcquinn2009, compostella2013}.

The basis for all these methods is the assumption that line widths are relatively directly related to the temperature of the absorbing gas. However, it is in fact well known that - in the simulations - the absorption lines are invariably wider than just the thermal broadening. In addition there is a very large {\em scatter} in line widths at a given density, in the sense that some lines are {\em much} wider than the thermal width at a given density.

In this paper we present a new model for measuring and interpreting
line widths, which builds on previous work. In our model lines
are {\em always} broader that the thermal width, and we also explain the
origin of the large scatter. We describe how the model can be used to
infer not just $T_0$ and $\gamma$, but also to constrain the size of
the absorbers. This paper is organised as follows. In Section~2 we begin by contrasting the \lq fluctuating Gunn-Peterson\rq\ description of the IGM with one where the absorption is due to a distribution of individual absorbers. We do so in order to establish notation and explain the motivation for our modelling. We discuss the properties of these absorbers, and in particular what sets their line widths. In Section~3 we introduce the simulations and use them to test the basic assumptions of the model. In Section~4 we examine whether the method can in principle be applied to observables.
We present a summary of our conclusions in Section~5.

\section{Lyman-$\alpha$ scattering in the IGM}
A Lyman-$\alpha$ photon has a large cross section for exciting a \ion{H}{I} atom from the ground state to the electronic $n=2$ level. 
When the atom falls back to the ground state, the photon is in general not re-emitted in the same direction. This scattering of Lyman-$\alpha$ photons out of the sight-line to a source is usually called Lyman-$\alpha$ \lq absorption\rq\, and is quantified by the optical depth $\tau$, such that $\exp(-\tau)$ is the resulting transmission - often referred to as the \lq flux\rq. We begin by calculating $\tau$ for the IGM in two very different approximations.

\subsection{The fluctuating Gunn-Peterson approximation}
The Lyman-$\alpha$ optical depth observed at wavelength $\lambda$ due to a {\em homogeneous} density distribution of neutral hydrogen with proper number density $n_{\rm HI}(z)$ in a cosmological setting, is given by 
\begin{equation}
\tau(\lambda)=\int_0^{z_s} \sigma \Big({\frac{(1+z)c}{\lambda}}\Big)\,n_{\rm HI}(z) \,\frac{c\,dz}{(1+z)H(z)}\,,
\label{eq:gp}
\end{equation}
\citep{gunn1965}, where $c$ is the light speed, $\sigma$ is the
Lyman-$\alpha$ cross section as function of frequency $c/\lambda$,
$\lambda_0\sim 1215.67$~{\AA} is the laboratory wavelength of the
\ion{H}{I} $n=1\rightarrow 2$ transition, and $H(z)$ the Hubble
parameter at redshift $z$ (see {\em e.g.} \citealt{meiksin2009} for an
extensive recent review of the relevant physics). Since the proper
length $dr$ corresponding to a redshift interval $dz$ is
$dr=c\,dz/\left[(1+z)H(z)\right]$, Eq.~(\ref{eq:gp}) simply states
that a column of neutral hydrogen $n_{\rm HI}dr$ contributes
$\sigma\,n_{\rm HI}dr$ to the optical depth. Here, $z_s$ is the
redshift of the source.

Structure formation results in the growth of the amplitude of fluctuations in the
density field. This causes the transmission as a function of wavelength to be in the form of relatively well defined \lq absorbers\rq\ \citep{lynds1971}, collectively referred to as the Lyman-$\alpha$ forest \citep{weymann1981}. At higher $z$ the absorbers blend into each other such that regions of near zero transmission are interrupted by small clearings (a Lyman-$\alpha$ jungle), whereas at low $z$ a near unity transmission is occasionally interrupted by a single absorber (a Lyman-$\alpha$ savannah). 

When the gas is in photo-ionisation equilibrium, the optical depth resulting from a single absorber depends on density and temperature as $\tau\propto \rho^2\,T^{-0.76}$ where $T^{-0.76}$ is (approximately) the temperature dependence of the hydrogen recombination rate. Taking peculiar velocities and thermal broadening into account as well results in the \lq fluctuating Gunn-Peterson\rq\ approximation \citep{gnedinhui1998,croft1998} for the optical depth,
\begin{eqnarray}
\tau(v) & = & \sum \int \frac{n_{\rm HI}}{1+z}\Big|\frac{dv'}{dx}\Big|^{-1} \sigma_{\alpha} dv' \label{eq:fulltau}\\
\sigma_{\alpha} & = & \sigma_0
\frac{c}{b\sqrt{\pi}}e^{-(v-v_0)^2/b^2} \nonumber\\
\sigma_0 &\equiv & \left(\pi e^2\over m_e c\right)\left(1\over4
  \pi\epsilon_0\right)f_{lu} \,,\\
\end{eqnarray}
where $e$ is the electric charge of an electron, $m_e$ the electron
mass, $\epsilon_0$ is the electric constant, and
$f_{lu}$ is the upward oscillator strength (usually referred to as the
\lq f\rq\ factor).  
The velocity coordinate $v$ is related to the observed frequency through
\begin{eqnarray}
\nu & = & \frac{\nu_0}{1 + \overline{z}}\left( 1- \frac{v}{c}\right)  \, ,
\end{eqnarray}
where $\nu_0=c/\lambda_0$ is the Lyman~$\alpha$ frequency in the lab,
$\overline{z}$ is a mean redshift of interest.  The integral is now over the profile of each individual \lq absorber\rq, the Jacobian $|dv'/dx|$ takes into account velocity gradients, and the sum is over many absorbers. The Gaussian profile expresses (thermal) broadening and neglects the intrinsic line profile of the Lyman-$\alpha$ transition. 

The temperature of the IGM is set by the balance between
photo-ionisation heating of the UV-background, the adiabatic compression,
and the adiabatic expansion of the universe. This introduces the
temperature-density relation which, for densities around the cosmic mean, is approximately a power-law (\cite{huignedin1997})
\begin{equation}
T=T_0\,\left({\rho\over \langle \rho\rangle}\right)^{\gamma-1}\equiv T_0\,\Delta^{\gamma-1}\,,
\label{eq:t-rho}
\end{equation}
where $T_0$ is the temperature at the mean density
$\langle\rho\rangle$, and $\Delta$ the density contrast. Close to reionisation the gas is nearly isothermal, $\gamma\sim 1$, whereas asymptotically long after reionisation, $\gamma\rightarrow 1+1/1.7$ \citep{huignedin1997,theuns1998}.
The $z<5$ Lyman-$\alpha$ forest can be reasonably well described as being a sum of of individual absorbers, and we discuss their properties next.

\subsection{Lyman-$\alpha$ absorbers}
The simplest single absorber, a cloud of uniform density, temperature $T$, and with H{\sc I} column density
$N_{\rm HI}$, produces a Gaussian absorption line when neglecting the intrinsic Lyman-$\alpha$ line profile,
\begin{eqnarray}
  \tau(v) & = & \tau_0 e^{-(v-v_0)^2/b^2} \nonumber\\
  \tau_0 & = & \frac{\sigma_0 c}{\sqrt{\pi}} \frac{N_{\rm HI}}{b}\,. 
  \label{eq:G1}
\end{eqnarray}
The line centre is at velocity
$v_0=c\log(1+z)+v_{\rm pec}$ for an absorber at redshift $z$, with
peculiar velocity $v_{\rm pec}$. Equation~(\ref{eq:G1}) can be obtained from Eq.~(\ref{eq:fulltau}) in the limit that $n_{\rm HI}|du/dx|^{-1}$ has a peak at $u_0$ with a width in velocity space much less than the thermal width $b$, this is the so called narrow-maximum limit in \cite{huignedinzhang1997}.

Several processes contribute to the broadening $b$ of the line. The thermal broadening is given by
\begin{eqnarray}
b_T^2&=& \frac{2 k_B T}{m_{\rm H}}\\
     &=& \left(12.8\,{\rm km~s}^{-1}\right)^2\,\,\left({\rm T_0\over 10^4~{\rm K}}\right)\,\Delta^{\gamma-1}\,\,,
\label{eq:btherm}
\end{eqnarray}
with $m_{\rm H}$ the mass of the hydrogen atom and $k_B$ Boltzmann's
constant. The second line applies to gas on a temperature-density relation of the form $T=T_0\,\Delta^{\gamma-1}$ from Eq.~(\ref{eq:t-rho}). 

A more realistic absorber should have at least {\em some} density
structure. Assuming the line centre corresponds to a maximum $\rho_0$
in density, a Taylor-expansion in velocity space yields to lowest
order
\begin{equation}
\log(\rho(v)/\rho_0) = - (v-v_0)^2/b_\rho^2\,.
\label{eq:Gr}
\end{equation}
Real Lyman-$\alpha$ absorbers may not be strongly peaked, but note
that what matters is not the density profile of the absorber, but its
{\em neutral} density. Since $n_{\rm HI}\propto \rho^2$ for isothermal gas
in ionisation equilibrium, peaks in neutral gas density are better defined
than the maxima in density - as we will show using numerical simulations in the next section.

The absorption profile of this more realistic absorber is then a
convolution of the Gaussians of Eqs.~(\ref{eq:G1}) and (\ref{eq:Gr}),
{\em i.e.} another Gaussian, with width
\begin{equation}
  b^2 = b_T^2 + b^2_\rho\,.
\label{eq:btot}
\end{equation}
What sets the velocity extent $b_\rho$? Absorbers at modest density contrast are expanding at a rate somewhat smaller than the Hubble rate, suggesting that $b_\rho\sim \lambda\,H$, with $\lambda$ the physical extent of the structure along the line of sight. Based on the papers of \cite{schaye2001} and \cite{gnedinhui1998}, we expect the physical extent of the absorbers to be of the order the Jeans length $\lambda_J$, defined in Eq.~(\ref{eq:lJ}) below \citep[see also][]{miralda1993} . In the model of \cite{schaye2001} absorbers are assumed to be locally in near-hydrostatic equilibrium, with column density related to density by
\begin{equation}
N_{\rm HI} \sim n_{\rm HI}\,\lambda_J\,,
\label{eq:statical}
\end{equation}
and hence \lq size\rq\ $\lambda_J$. Because the dynamical times of low-density absorbers are long, the structures are still expanding, and are not exactly in hydrostatic equilibrium but are lagging behind. This is encapsulated by the \lq filtering
scale\rq\ $\lambda_F=2\pi/k_F$ of \cite{gnedinhui1998} which depends
on the full history of the Jeans length over cosmic time and hence depends on the thermal history, not just the instantaneous Jeans length. 
The physical \lq extent\rq\ of the absorber in this approximation is then $\sim 1/k_F$, therefore its velocity extent is $\sim H(z)/(k_F\,(1+z)) \approx \lambda_J\,H(z)/(2\pi)$, where $\lambda_J$ is now the physical - as opposed to co-moving - extent of the absorber. This then motivates us to write
\begin{equation}
b_\rho = f_J\,{\lambda_J\,H(z)\over 2\pi}\,.
\label{eq:Gd}
\end{equation}
The factor $f_J$ parametrises the time-dependent Jeans smoothing of the gas density profiles, and we expect it to be of order unity. 

For gas on a power-law temperature-density, the values of the sound
speed and Jeans length depend on the density contrast $\Delta$ as
\begin{eqnarray}
c_s^2 &=& {\Big( \frac{5 k_{\rm B}T}{3\mu m_{\rm H}}\Big)}\\
       &= &(15.2\,{\rm km~s}^{-1})^2\,\left({T_0\over 10^4~{\rm K}}\right)\,\Delta ^{\gamma-1}\,\nonumber\\
\lambda_J^2 &\equiv & {c_s^2\pi\over {\rm G}\rho_{\rm t}}\nonumber\\ 
            &=& {(5/3)\pi k_{\rm B}\,T_0\over \mu m_{\rm H}\,{\rm G} \rho_t(z=0)\,(1+z)^3}\,\Delta^{\gamma-2}\nonumber\\ 
            &\approx &\left(260\,{\rm kpc}\right)^2\,\left({T_0\over
                10^4~{\rm K}}\right)\,\Delta^{\gamma-2} 
            \nonumber\\
            &\times&\left({\Omega_m\over 0.307}\right)^{-1}
            \left({h\over 0.6777}\right)^{-2}\left({1+z\over 4}\right)^{-3}\,,
\label{eq:lJ}
\end{eqnarray}
where $\mu$ is the mean molecular weight and $\rho_{\rm t}=\rho_{\rm
  dm}+\rho_{\rm b}$ is the {\em total} density, dark matter plus
baryons, $G$ is the gravitational constant, $\Omega_m$ is the matter
density. We write the Hubble parameter at redshift $z$ as $H(z)$, with $H_0=100h$~km~s$^{-1}$~Mpc$^{-1}$ its value today. The numerical values assume $\mu=0.59$, that dark matter and gas move together, and for reference we use the \cite{Planck13} values for the cosmological parameters. Numerically, the broadening due to Jeans smoothing is thus of order
\begin{eqnarray}
b_\rho^2 &=& f_J^2\,{10\,k_{\rm B}T\over 9\mu\,m_{\rm H}\Delta}\,{H^2(z)\over H^2_0\,\Omega_m\,(1+z)^3}\\
&\approx &
 (11.1\,{\rm km~s}^{-1})^2\,
 \left({f_J\over 0.88}\right)^2\,\left({T_0\over 10^4~{\rm
       K}}\right)\,\Delta^{\gamma-2}\,\,, \nonumber
       \label{eq:brho}
\end{eqnarray}
independent of redshift in the high-redshift approximation
$H(z)\propto (1+z)^{3/2}$. The
model of \cite{hui1999} uses a very similar reasoning to infer line
widths; $b_\rho$ is usually referred to as the \lq Hubble
broadening\rq. We demonstrate below that a value of $f_J\sim 0.88$
describes our numerical simulations well at $z=3$.

The total broadening, $b$, of the absorber is then
\begin{eqnarray}
b^2 &=& b_T^2\,\left[1 + f_J^2\,{5\over 9\mu}\,\Delta^{-1}\right]\nonumber\\
    &\approx & b_T^2\,\left[1 + 0.75\left({f_J\over 0.88}\right)^2\,\Delta^{-1}\right]\,.
    \label{eq:bfinal}
\end{eqnarray}
The line-widths of weak absorbers ($\Delta\lsim 1$) are increasingly
dominated by Hubble broadening, $b^2\sim
b_T^2\,\Delta^{-1}\propto\Delta^{\gamma-2}$, and hence the filtering
scales determines their widths. Hubble expansion plays  a smaller role
for stronger lines, with line widths ultimately set by thermal
broadening alone. Equation~(\ref{eq:bfinal}) smoothly interpolates
between filtering and pure thermal broadening. However, we note that
for typical $z\sim 3$ absorbers with $\Delta\sim 3$, Hubble broadening
still plays a significant role, increasing $b$ by almost 70 per cent
compared to thermal broadening alone. It might be possible to measure
the extent of absorbers in real space using close quasar pairs (see
\cite{rorai2013}; the impact of Jeans length on quasar pairs was also
considered in \cite{peeples2010II}). The extra broadening discussed here is a longitudinal measure of the Jeans length, as opposed to the transverse Jeans length probed with pairs.

The Gaussian shape of the absorber directly relates column density and width,
\begin{eqnarray}
b^2 &=& -{2\tau^2\over \tau\,\tau'' - \tau'^2} = -\frac{2
  \tau_0}{\tau_0''}\,
\label{eq:b}\\
N_{\rm HI} &=& \frac{\sqrt{\pi}}{\sigma_0 c} b \tau_0\,,
\label{eq:nhi}
\end{eqnarray}
where $\tau'\equiv d\tau/dv$, $\tau''\equiv d^2\tau/dv^2$, and the
subscript 0 indicates that the quantity is to be evaluated at the
(local) maximum of the optical depth. We will use these relations
below. Equations~(\ref{eq:b}-\ref{eq:nhi}) follow directly from Eqs.~(\ref{eq:G1}).

\section{Broadening of Lyman-$\alpha$ absorbers in simulations}
\subsection{Mock spectra from simulations}
\begin{figure}
  \centering
  \subfloat[]{\includegraphics[width=\columnwidth]{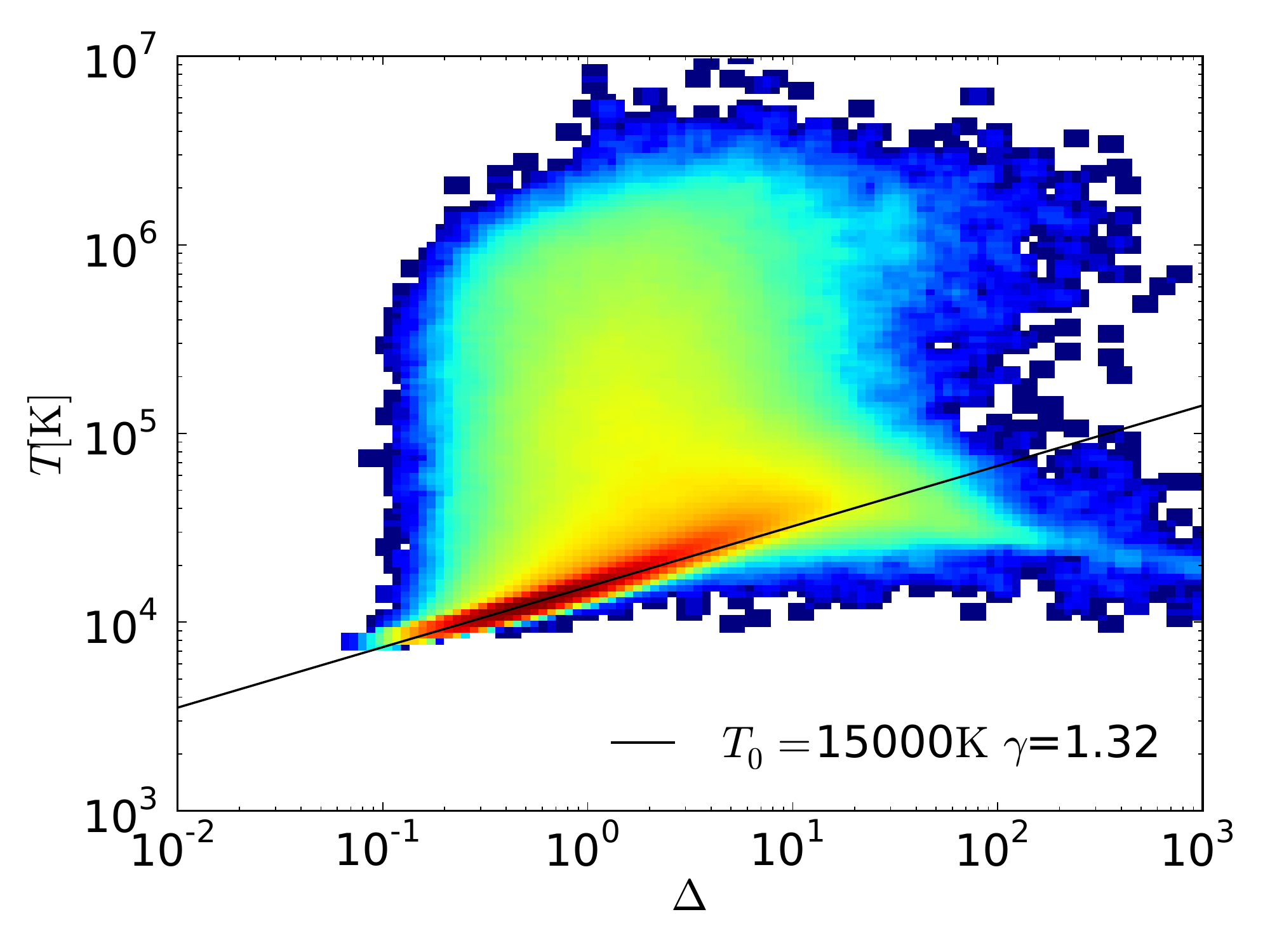}\label{fig:tdelta}}
  \newline
  \subfloat[]{\includegraphics[width=\columnwidth]{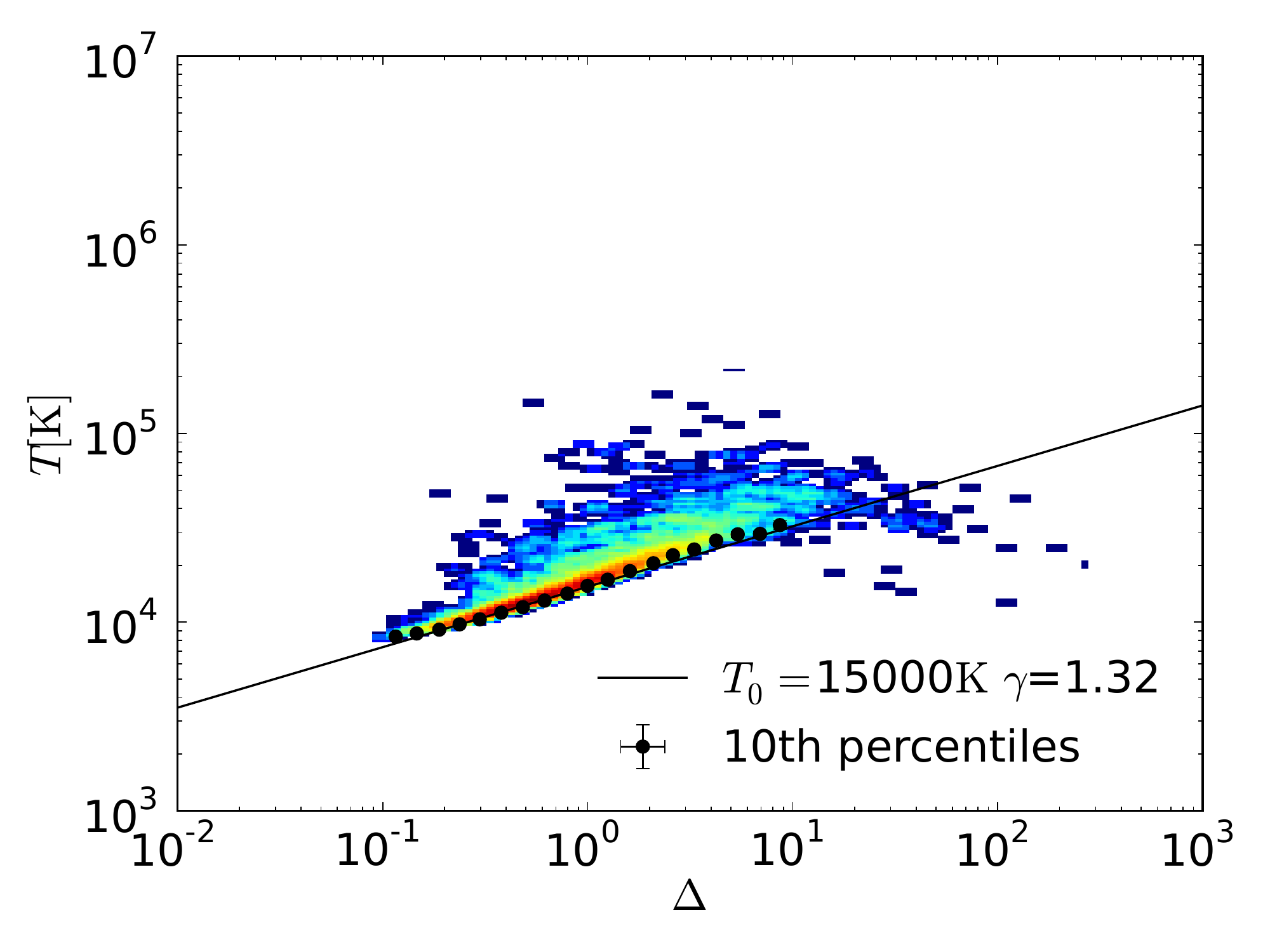}\label{fig:tdelta_lines}}
  \caption{Mass-weighted (top panel) and optical depth weighted (bottom panel) temperature-density relation 
  in the {\sc reference} {\sc owls} run at $z=3$. Values in the bottom
  panel are evaluated at the locations of maxima in the optical
  depth. Colours are a measure of the density of points in these plots, and
  $\Delta=\rho/\langle\rho\rangle$. In the lower panel the black dots indicate the median
    temperature in bins of $\Delta$ with error bars computed by bootstrap resampling. The black solid line
    in both panels show the power-law $T=T_0\Delta^{\gamma-1}$ that best fits
    the black dots for $\Delta <3$. There is a much larger scatter in $T$ vs $\Delta$ in the top panel than in the bottom panel, but this shocked gas contributes little to the Lyman-$\alpha$ forest. The lower-envelope of points in the $T-\Delta$ relation
    in the bottom panel mirrors the temperature-density relation of
    the lower-envelope in the top panel, demonstrating that the
    optical-depth weighted temperature at the optical depth maxima of
    absorption lines follows very closely the $T-\Delta$ relation of
    the photo-heated IGM.}
\end{figure}
\begin{figure*}
  \centering
  \includegraphics[width=0.9\textwidth]{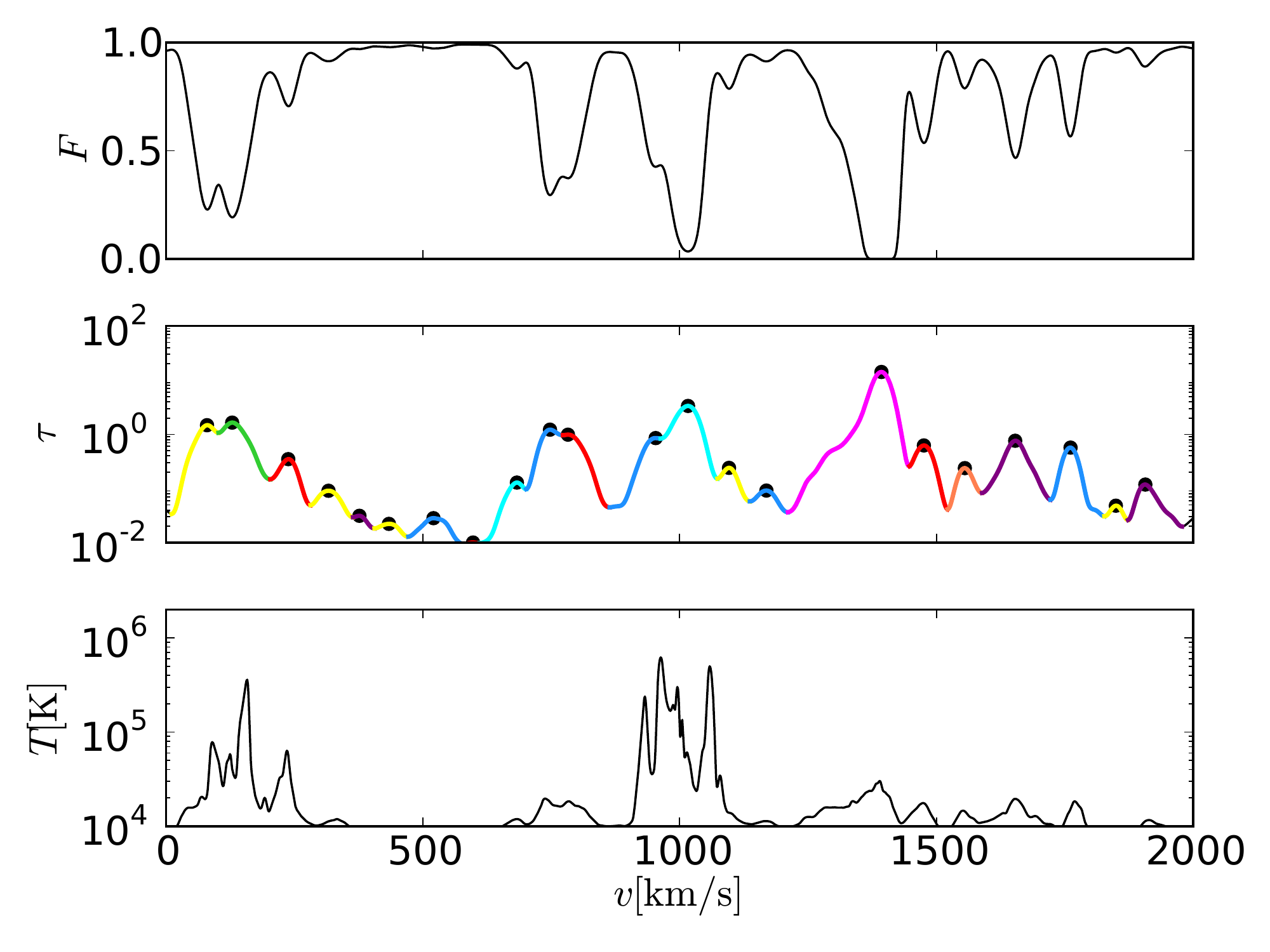}
  \caption{A stretch of a simulated spectrum at redshift
    $z\sim3$. From top to bottom we show the
    transmission $F\equiv(-\tau)$, optical depth $\tau$, and optical
    depth weighted temperature, $T$, as function of velocity
    $v$. Individual lines are regions of the spectra between two minima
    in optical depth, they are coloured using different colours, with a
    filled circle at the maximum $\tau$ for each line. Note that $T$ is not constant over a line.}
  \label{fig:spectra}
\end{figure*}

We use the {\sc reference} model of the {\sc owls} suite \citep{schaye2010} to test
our model for line broadening. Briefly, the {\sc owls} suite is a set 
of cosmological hydrodynamical simulations performed with the {\sc gadget-3} code,
an improved version of {\sc gadget-2} last described by \cite{springel2005}, with subgrid models for unresolved
galaxy formation physics, in particular star formation, and feedback by and enrichment from stars,
as described in \cite{schaye2008}, \cite{dallavecchia2008} and \cite{wiersma2009b}, respectively.
Although efficient feedback from star formation results in outflows from galaxies in this simulation, such winds
do not have a significant impact on the Lyman-$\alpha$ forest for reasons discussed in \cite{theuns2002c}, see also \cite{viel2013}.  Most relevant for this paper is that we
simulate a $\Lambda$ cold dark matter model using 512$^3$ SPH particles that represent the gas, and an equal number of dark
matter particles. The computational volume is a periodic box of co-moving size 25$h^{-1}$~Mpc on
a side. The masses of the gas particles ($\sim 2\times 10^6 h^{-1} M_\odot$) are sufficiently low to obtain numerical convergence for the line widths \citep{theuns1998}. Photo-heating of the gas in the presence of
an optically thin UV/X-ray background from \cite{haardt2001} is
implemented using interpolation tables from \cite{wiersma2009a}.

Photo-heating, adiabatic processes, shocks from
structure formation and feedback from galactic winds and radiative cooling produce the temperature-density relation shown in 
Fig.~\ref{fig:tdelta} at redshift $z=3$. Most of the gas is on a tight temperature-density relation, $T=T_0\,\Delta^{\gamma-1}$, with $T_0\approx 15000$~K and $\gamma\approx 1.32$ for $\Delta\lsim 5$, and this gas 
is the main reservoir responsible for producing the Lyman-$\alpha$ forest. There is gas at significantly higher temperatures at these densities, which results from structure formation and feedback. The increased importance of radiative cooling decreases the amount of hot gas at higher densities, and an equilibrium sequence where radiative cooling and photo-heating balance results in the appearance of significant amounts of gas at $T\sim 10^4~$K. This higher density gas results in Lyman-limit and damped Lyman-$\alpha$ absorbers (e.g.~\citealt{altay11}). Self-shielding is important for this higher-density gas, and the impact of feedback becomes more important \citep{altay13a}.

Given a snapshot of the simulation at some redshift (we use $z=3$ in
this paper), we generate mock Lyman-$\alpha$ spectra using the method
described in \cite{theuns1998} but using the interpolation tables of
\cite{wiersma2009a} to relate total to neutral hydrogen density. We
generate these spectra at very high resolution using pixels of
width $\sim 2$~km~s$^{-1}$, much narrower than any of the absorption
features that appear. Importantly, we neither add noise nor
instrumental broadening to mimic observed spectra - limiting our analysis to comparing our model to idealised observations.
Each sight line is parallel to a coordinate axis of the simulation
volume, and we use velocity
pixels of size $v=H(z)L/[(1+z)N]$, where $L=25h^{-1}$~Mpc is the
co-moving box size, and $N$ the number of pixels. We calculate $\tau(v)$, $T(v)$ and $\rho(v)$,
where $T$ and $\rho$ are the {\em optical depth weighted} temperature
and density, respectively (see~\citealt{schaye1999}). Unless explicitly stated, we will always use optical
depth weighted quantities for variables measured from spectra.

In Fig.~\ref{fig:tdelta_lines} we plot values of optical depth
weighted temperature versus density, for pixels that correspond to
local maxima in the optical depth from these mock spectra. Most points
lie close to the lower-envelope of the points in the scatter plot, which tracks the same $T-\Delta$ relation as the lower envelope in Fig.~\ref{fig:tdelta}. The scatter around this relation is much less in Fig.~\ref{fig:tdelta_lines}, because the shocked gas that causes the scatter to higher $T$ in Fig.~\ref{fig:tdelta} contributes little to the optical depth. At higher $\Delta\gsim 10$, $T$ decreases with increasing $\Delta$, reflecting the effects of radiative cooling also apparent in Fig.~\ref{fig:tdelta}.

\subsection{Identifying individual absorbers}
We identify individual absorbers (or \lq lines\rq) with regions of the spectrum between two
minima in the optical depth, dissecting the whole spectrum in a set of
unique stretches. An example spectrum is shown in
Fig.~\ref{fig:spectra}, with different lines coloured differently, and
with diamonds at the locations of local maxima in $\tau$. Unsurprisingly, the temperature varies across a line, and
hence assigning a single temperature (or density) to a line is
necessarily somewhat ambiguous.  We could decide to associate
to a line the maximum, the (weighted) mean or the value corresponding to the maximum
of the line for any given physical quantity. In the following, where not
specified otherwise, we will associate to the line the value of the
physical quantity that corresponds to the optical depth weighted value evaluated 
at the local maximum in optical depth. This method for identifying
lines is not directly applicable to observed spectra in the presence
of noise. Such spectra should be smoothed to avoid incorrectly
identifying noise features with small lines, however this may lead to
missing true weak absorbers. Preliminary analysis shows that fitting a
Gaussian over a set of contiguous pixels to calculate the derivatives
that appear in Eq.~(\ref{eq:b}) is promising, we will report on this
in a future paper.

\begin{figure}
  \centering \includegraphics[width=\columnwidth]{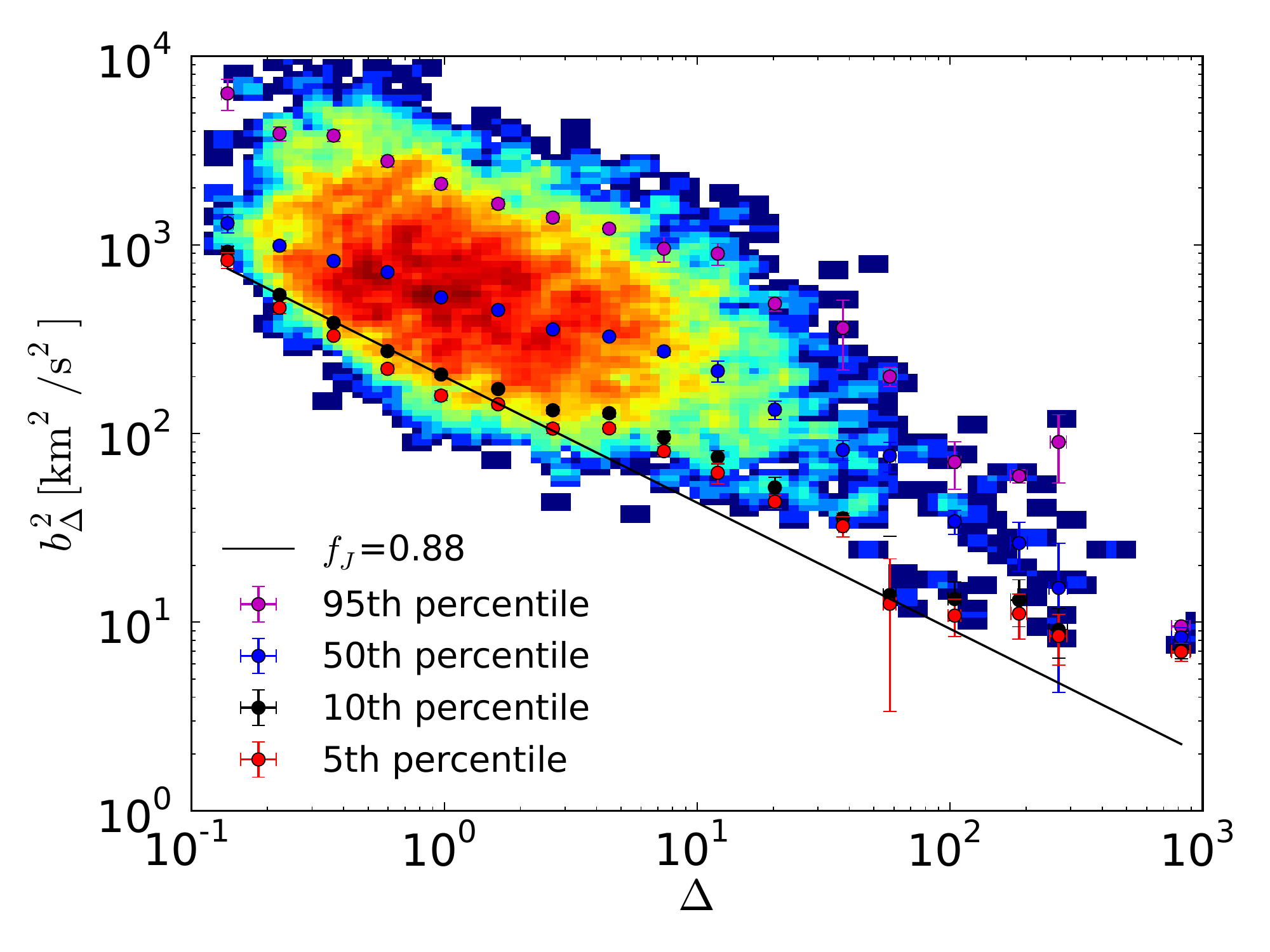}
  \caption{Velocity extent of absorbers in real space $b_\Delta$
    versus density contrast $\Delta$ for lines identified in mock spectra without peculiar velocities. Colours
  indicate density of points in the plot.
   The 5th, 10th, 90th and 95th percentiles of $b_\Delta$ in bins of $\Delta$ are shown as symbols with bootstrap errors. Line widths
    {\em decrease} with increasing density, 
    opposite to the case of thermal broadening.
    This behaviour is captured by  
    the solid black line, the broadening $b_\rho$ from  Eq.~(\ref{eq:Gd}) for $f_J=0.88$, which fits the lower percentiles of the $b_\Delta-\Delta$ relation well in shape and amplitude for $\Delta\lsim 3$.
    }
 \label{fig:b2hubble}
\end{figure}

\begin{figure}
  \centering \includegraphics[width=\columnwidth]{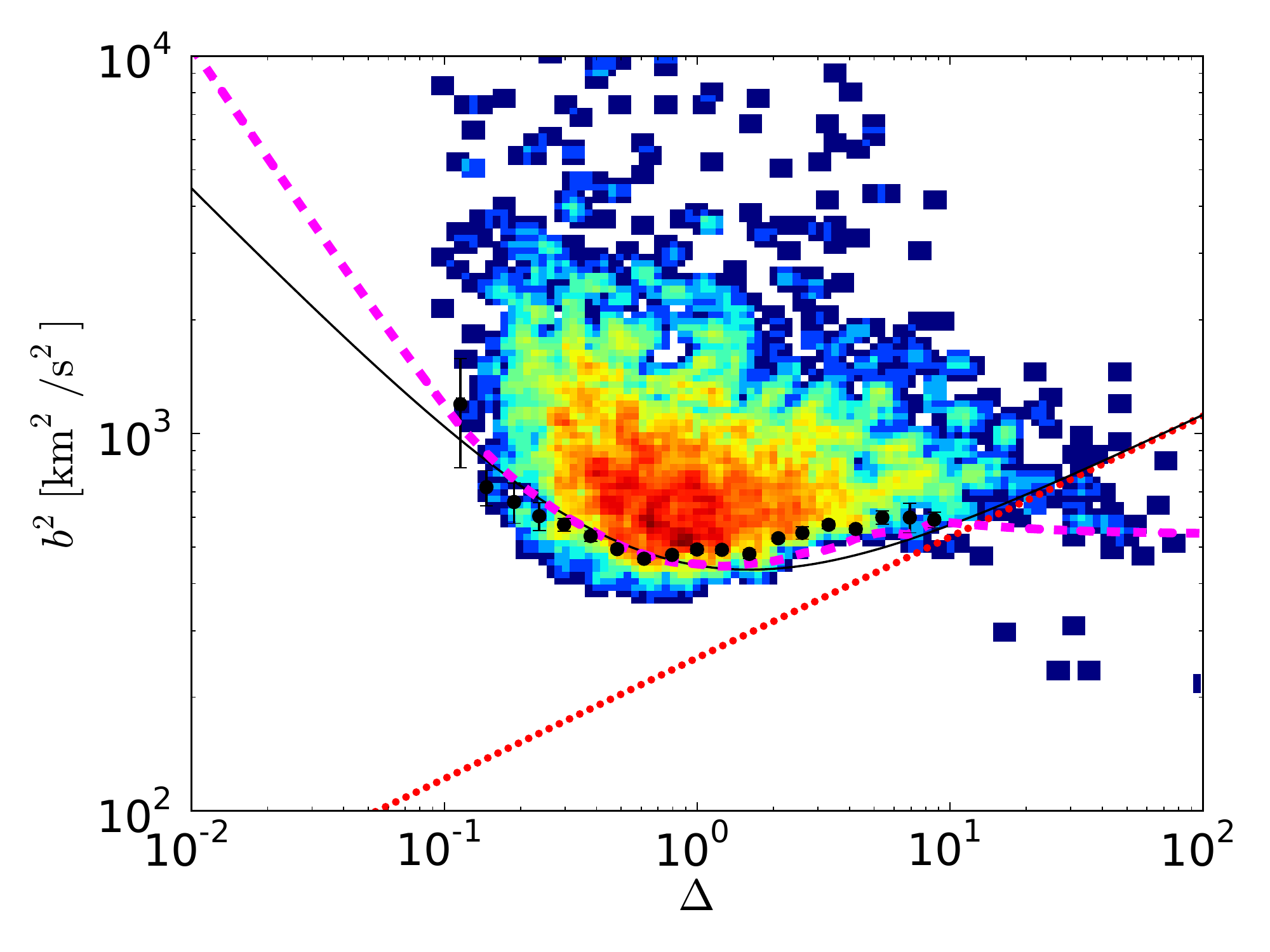}
  \caption{Total broadening $b^2\equiv -2\tau/\tau''$ of absorbers at line centres as a function of density
    contrast $\Delta$ for lines identified in mock spectra. Colours
    indicate density of points in the plot. {\em Black dots} with bootstrap error bars
    denote the location of the 10th lower percentile
    of $b$ in bins of $\Delta$.     
    The {\em dotted
      red} line is the thermal width from Eq.~\ref{eq:btherm} for the temperature-density relation 
      from Fig.~\ref{fig:tdelta}; lines are invariably wider
    than $b_T$ except at $\Delta\gsim 10$. The {\em solid black line} is the model of
    Eq.~(\ref{eq:btot}) with $f_J=0.88$, for the same $T-\Delta$ power-law relation.
     It describes the black dots very well up to
    $\Delta\sim 3$, above which the simulation's $T-\Delta$ relation is
    not well fit by a power-law. The {\em dashed magenta line} is again the model of Eq.~(\ref{eq:btot}) but now using the measured $T-\Delta$ relation rather than a power-law fit: it captures the down-turn in $b$ for $\Delta\gsim 3$ due to the onset of radiative cooling.}
  \label{fig:broadening}
\end{figure}

\begin{figure}
  \centering \includegraphics[width=\columnwidth]{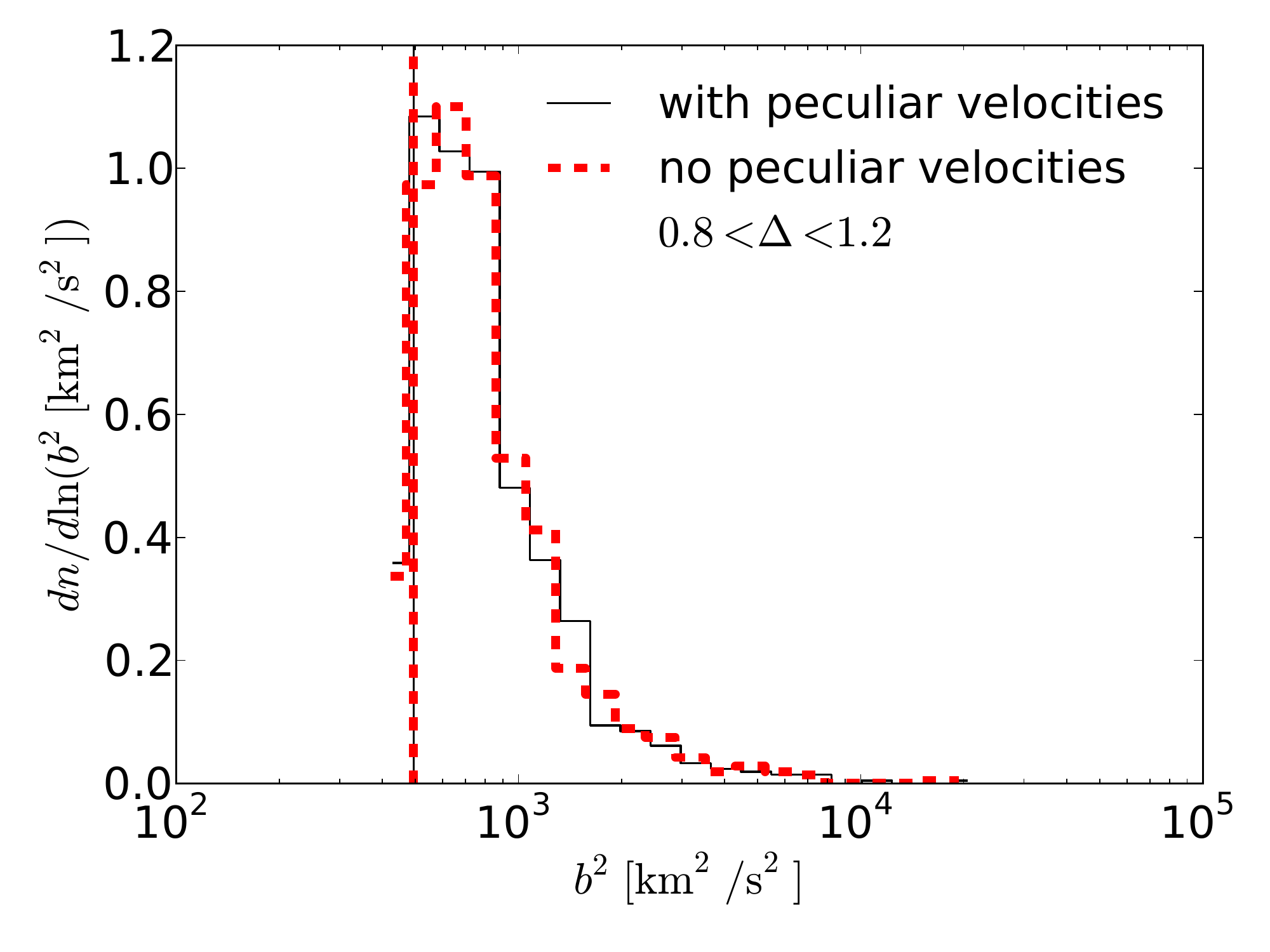}
  \caption{Histogram of line broadening $b^2=-2\tau/\tau''$ for
    simulations with and without peculiar velocities, shown as {\em
      solid black} and {\em dashed red} lines, respectively.  The 10th percentiles
    of narrowest lines are indicated by vertical lines. The
    impact of peculiar velocities on the distribution of $b$ values is
    small.}
  \label{fig:pecvel}
\end{figure}

As demonstrated in Fig.~1, the (real-space)
temperature-density relation of the IGM in the simulation translates
approximately to a power-law (optical depth weighted)
temperature-density relation for the lines generated in mock spectra.
Most lines follow $T=T_0\propto\Delta^{\gamma-1}$ up to $\Delta\lsim
3$, with some scatter to significantly higher $T$, but none scatter
significantly {\em below} this line. However for $\Delta\gsim 3$, the
median temperature does fall below the power law fit: this is because
this gas cools radiatively, and this results in a decrease in $T$ with
higher $\Delta$ for the real-space temperature-density relation as
well. Note further that the $T-\Delta$ relation is only approximately
a power-law, with evidence for some curvature even over a relatively
narrow $\Delta$ range.  Finally, note that in the absence of noise, we
see that the majority of lines have $0.1\lsim \Delta\lsim 10$.

The optical depth profile of several of the lines in Fig.~\ref{fig:spectra} appear reasonably Gaussian in shape near their maxima, and we will investigate to what extent their widths follow Eq.~(\ref{eq:bfinal}). However it is also clear that some lines  blend with nearby lines, and we will show below that this \lq clustering\rq\ impacts how well we can infer the underlying $T-\Delta$ relation from spectra - even in the absence of noise.

We suggested that line broadening has a contribution $b_\rho$ due to
the spatial extent of the absorber as described by
Eq.~(\ref{eq:Gd}). To test this directly, we should plot $b_\rho$
versus $\Delta$. However, maxima in $\tau$ do not correspond directly
to maxima in $\Delta$ because of peculiar velocities. Therefore, to
unambiguously identify a line with a density structure, we generate
mock spectra setting peculiar velocities to zero. We later show that
this has very little effect on the line broadening (see also
Fig.~\ref{fig:pecvel}). For each line identified in these mock
spectra, we can now unambiguously identify the corresponding density
stretch that gives rise to that line. However, the shape of the
density structure of lines is often complex and it is not immediately
apparent what extent to associate with a given line. We decided to use
the following prescription. For each line we determine its start and
end velocity, $v_1$ and $v_2$, the location ($v_{\rm max}$) and height
($\Delta_{\rm max}$) of the maximum density, and the integral
$I\equiv\int_{v_1}^{v_2} \Delta(v)\, dv$. We now determine the value
of $b_\Delta$ of the Gaussian profile, $G(\Delta)=\Delta_{\rm
  max}\,\exp(-(v-v_{\rm max})^2/b_\Delta^2)$, for which
$\int_{v_1}^{v_2} G(v)\, dv=I$. In other words, we define the extent
of the absorber, $b_\Delta$, as the width of the Gaussian that has the
same integral and maximum value as the line itself. If the line {\em
  were} a Gaussian, then $b_\Delta$ would simply be its width.
Further details on the fitting procedure can be found in
Appendix~\ref{app:fj}. We expect that $b_\Delta\sim b_\rho$ from
Eq.~(\ref{eq:Gd}).

In Fig.~\ref{fig:b2hubble} we plot $b^2_\Delta$ versus $\Delta$ for
all lines identified in mock spectra generated while ignoring peculiar
velocities. As before, colour encodes the number density of lines
in this plot, and 5th, 10th, 50th and 95th percentiles are plotted as
symbols with bootstrap error bars. Values of $b^2_\Delta$ {\em
  increase} with decreasing $\Delta$ - the opposite of what is
expected from thermal broadening. The solid line indicates $b^2_\rho$
from Eq.~(\ref{eq:Gd}) with $f_J=0.88$ applied to gas with the
$T-\Delta$ relation $T=15000~{\rm K}\,\Delta^{1.3-1}$ which, as we
showed earlier, fits  the actual temperature-density relation of the
simulation well for $0.1\lsim\Delta\lsim 3$. Fig.~\ref{fig:b2hubble} shows that both the 5th and 10th percentiles follow the solid line well over the same $\Delta$ range. This implies that $b_\rho$, with $f_J=0.88$, is indeed a good estimate of the width of the absorbers in real space. 

\begin{figure*}
  \centering \includegraphics[width=0.9\textwidth]{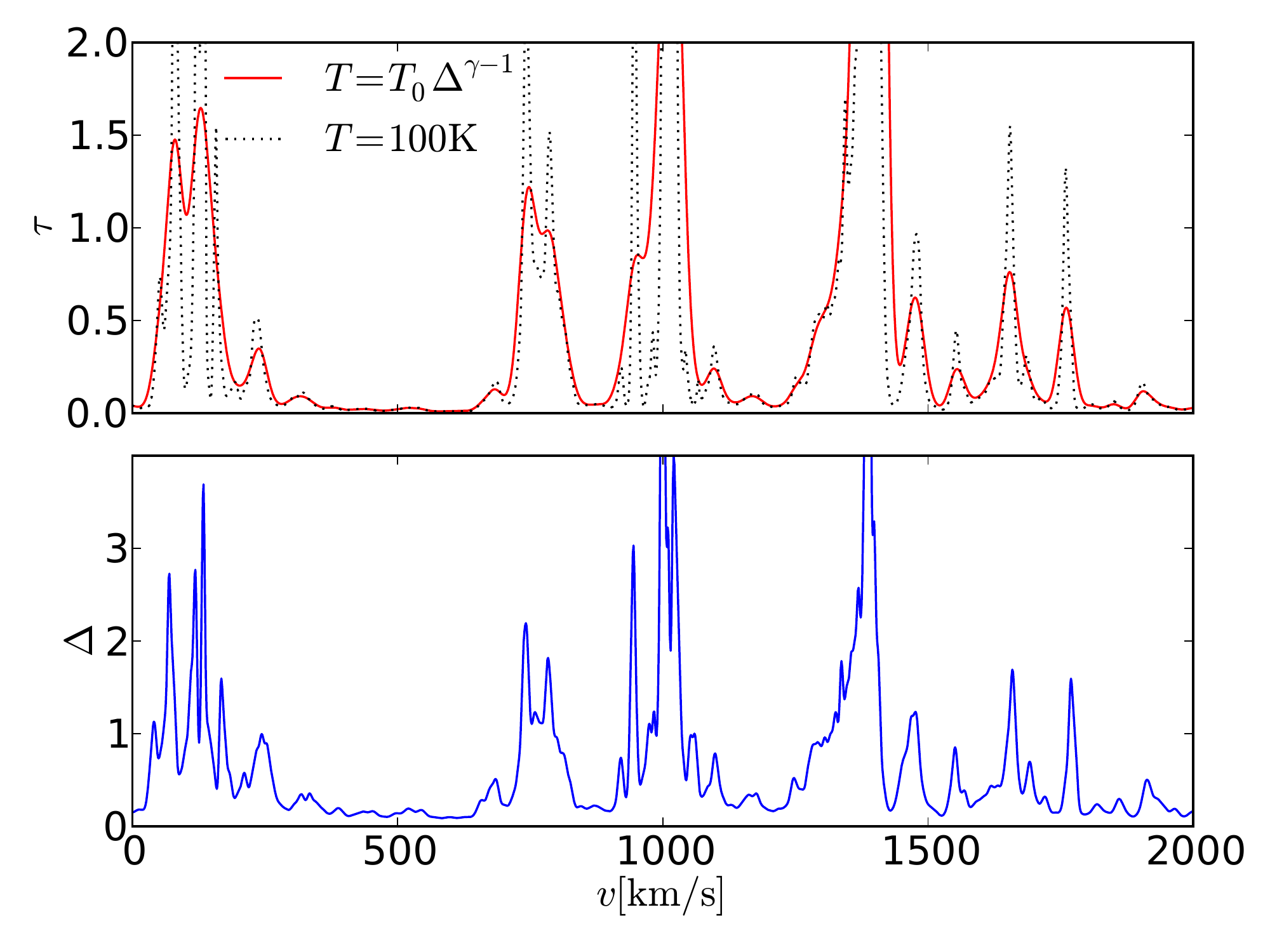}
  \caption{Optical depth $\tau$ ({\em top panel}) and corresponding
    density contrast $\Delta$ ({\em bottom panel})
  for mock spectra for the simulation with the original
  temperature-density relation ({\em solid red line})
  and for the case where $T=100$~K everywhere ({\rm dotted black line}). Clustering of density peaks in real space makes some of the
  absorption lines much wider than the estimate of $b$ from Eq.~(\ref{eq:bfinal}), for example the wider lines around $v\sim 800$~km~s$^{-1}$ and $v\sim 1000$~km~s$^{-1}$. Thermal broadening and Hubble broadening together smooth the individual peaks seen in the underlying density field into one wide line, whose width cannot be described accurately by eq.~\ref{eq:bfinal}.}
  \label{fig:clustering}
\end{figure*}

Given that our estimate of the broadening of the density profile works
well, we plot the total line broadening (including peculiar velocities) versus density contrast, and compare it to the model of Eq.~(\ref{eq:bfinal}) (Fig.~\ref{fig:broadening}).
As before, the colour represents the
density of points in the plot. We compute $b$ from the curvature of
the optical depth profile $\tau(v)$ using Eq.~(\ref{eq:b}); fitting a
Gaussian to the pixels close to the maximum yields very similar values
of $b$. Values of $b-\Delta$ for individual lines show a larger
scatter in $b$ at given $\Delta$, but there is a well-defined lower
envelope below which few lines appear. This lower envelope is traced
by the 10th percentile of $b$ values in narrow bins of $\Delta$,
plotted using black symbols with bootstrap errors. The red dashed line
indicates the thermal broadening, $b_T$, from Eq.~(\ref{eq:btherm}),
using the power-law fit to the temperature-density relation measured
in the simulation. For $\Delta\lsim 10$, the measured broadening along
the cut-off is significantly wider than $b_T$, with the discrepancy
between the two increasing with decreasing  $\Delta$. The solid black
line shows $b_T^2+b_\rho^2$ as a function of $\Delta$ from
Eq.~(\ref{eq:bfinal}), again using the best-fitting power-law
temperature-density relation, and adopting the value $f_J=0.88$ found
above. It captures accurately the dependence of the broadening on
$\Delta$ for the lower envelope of the absorption lines traced by the
10th percentile for $\Delta \lsim 3$, in particular reproducing the
measured {\em upturn} in $b^2$ for $\Delta \lsim 1$. Because the solid
black line assumes that the relation between temperature and density
is a power law, it does not describe the simulated $b-\Delta$ relation
well above $\Delta\sim 3$. The dashed black line uses the median value of the temperature at a given density for absorption lines when calculating the Jeans smoothing term $b_\rho$ in Eq.(\ref{eq:Gd}), rather than a power-law. It does much better in capturing the downturn in $b-\Delta$ at $\Delta\gsim 10$. We conclude from this plot that the model in which the line broadening is a combination of thermal and Jeans smoothing,
Eq.~(\ref{eq:bfinal}) describes the lower-envelope in the $b-\Delta$
plane well. Note that deviations of the true temperature-density
relation from a power-law, and Jeans smoothing, cause the $T-\Delta$
relation to be non-monotonic, with a minimum value of $b$ for
$\Delta\sim 1$ in this particular simulation at this particular redshift.

Peculiar velocities are not the cause of the appearance of lines that are much {\em wider} than the broadening $b$ computed from
Eq.~(\ref{eq:bfinal}). We compare the distribution of $b$ values for lines identified in mock spectra with and without peculiar velocities in Fig.~\ref{fig:pecvel}. The peculiar velocities do not have
a large effect on the broadening, if anything they make lines slightly {\em narrower}, see also \cite{theuns2000}. 

Our analytic expression for broadening, Eq.~(\ref{eq:bfinal}),
provides only a lower limit on the width of a collection of lines in
the Lyman~$\alpha$ forest. The additional broadening can be partially
attributed to the underlying density
structure in the absorber \citep{hui1999}, for example in terms of the
angle under which the sightline intersects the filament that
corresponds to the absorber. However, another important contribution
that is not easily captured by linear theory is the {\em clustering} of
density peaks.  We illustrate  this in Fig.~\ref{fig:clustering},
comparing a mock spectrum with the original $T-\Delta$ relation with
the same spectrum now computed assuming $T=100$~K everywhere. We see
that some individual absorption lines identified in the solid red
spectrum have several underlying density maxima. If the temperature of
the gas is low, as in the case of the black dotted spectrum, then
these individual density maxima give rise to their own absorption
lines. But if the gas is sufficiently hot, then thermal and Hubble
broadening can merge multiple individual components into one much wider line. The width of that absorber is not described well by our model. We conclude that it is the clustering of density maxima that gives rise to the lines that are much wider than expression $b$ from Eq.~(\ref{eq:bfinal}). Another way to
demonstrate this is to count the number of individual density peaks for each absorber: we find that absorbers with widths close to 
that from Eq.~(\ref{eq:bfinal}) typically have only one density maximum, whereas those that are significantly wider correspond to several underlying peaks.

\section{Inferring the temperature-density relation from observables}
\begin{figure}
  \centering 
\includegraphics[width=0.9\columnwidth]{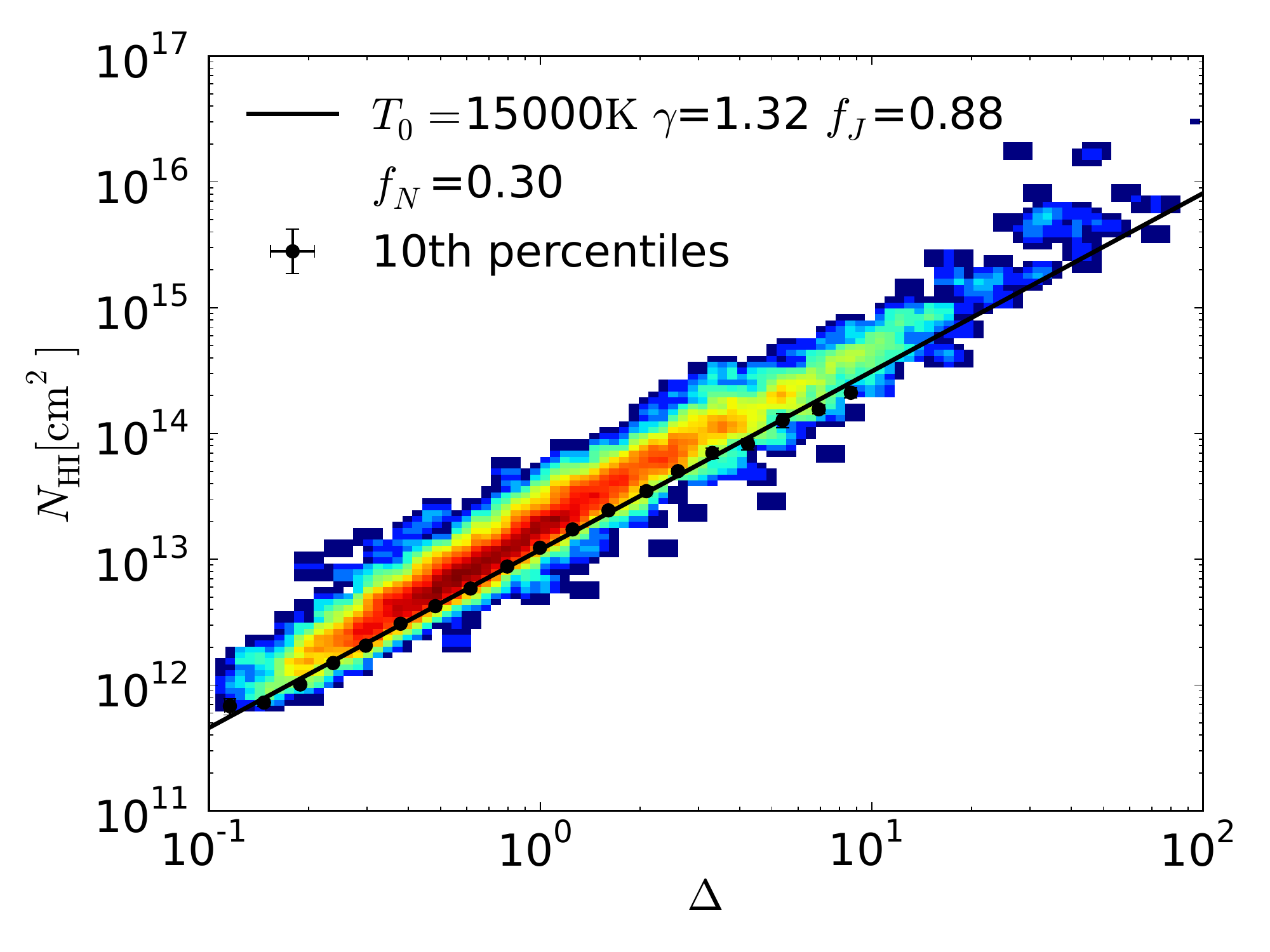}
\caption{The neutral hydrogen column density of absorption lines
  versus density contrast for the {\sc reference} model at $z=3$;
  colours are a measure of the density of points in the plot. {\em
    Black dots} denote the 10th percentiles of column density in
  logarithmic bins of $\Delta$. The {\em solid black line} corresponds
  to the model of Eq.~(\ref{eq:nhi}), using the value of $\Gamma$
  appropriate for the {\sc reference} model with values of $T_0$,
  $\gamma$ and $f_J$ taken from Fig.~\ref{fig:broadening}, and for
  $f_N=0.3$. This value of $f_N$ is the result of best-fitting
  Eq.~(\ref{eq:nhi}) to the 10th percentiles, for $\Delta\leq
  0.3$. This provides an excellent fit to the results from the
  simulation over 4 orders of magnitude in column density.  }
  \label{fig:nhi_rec_delta}
\end{figure}

\begin{figure}
  \centering \includegraphics[width=\columnwidth]{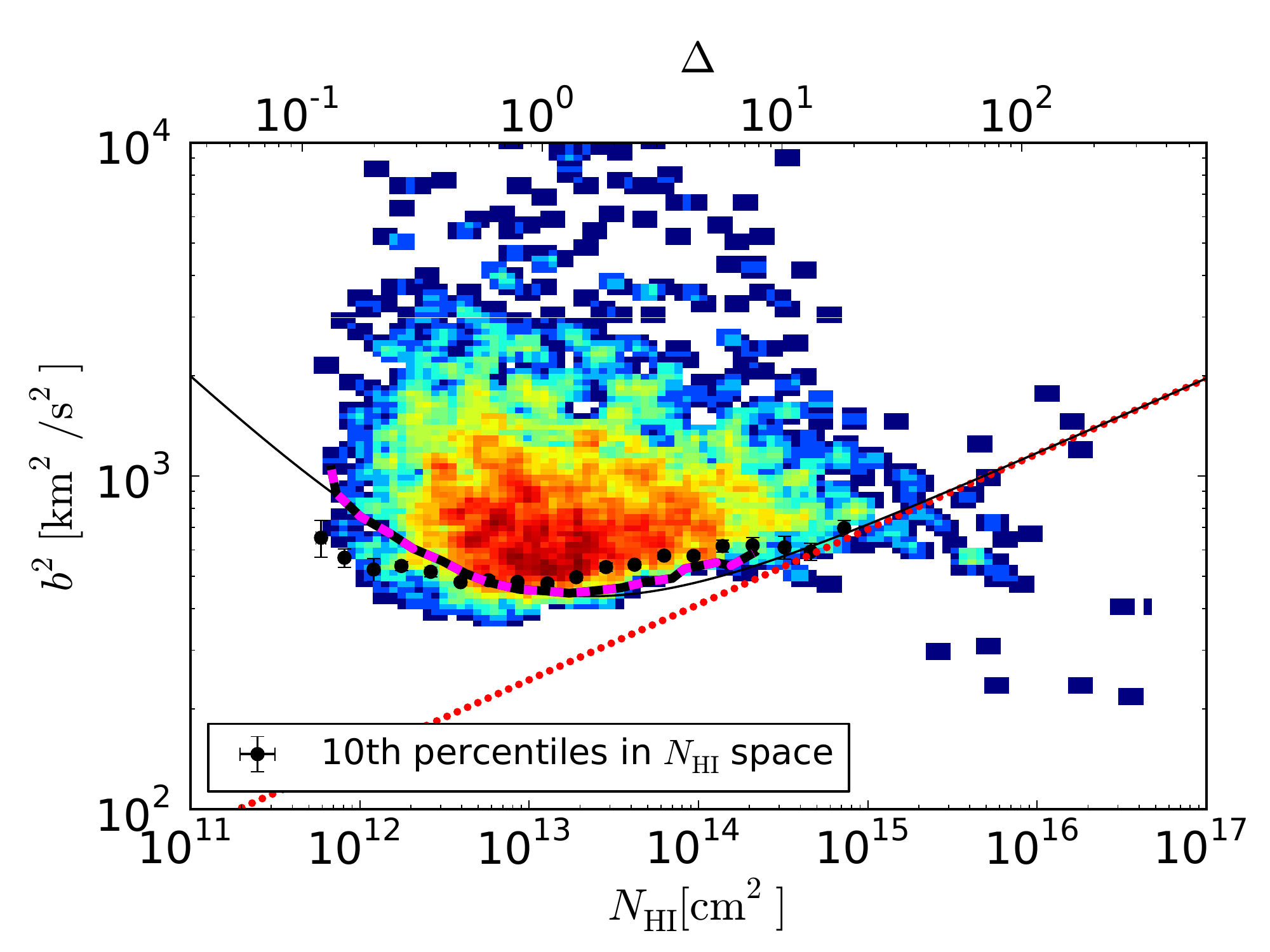}
  \caption{The line-width column density $b^2-N_{\rm HI}$ relation in
    the {\sc reference} {\sc owls} model at redshift $z=3$, with
    broadening determined from Eq.~(\ref{eq:b}) and column density
    from Eq.~(\ref{eq:nhi}). The top axis gives the value of the
    density contrast that corresponds to $N_{\rm HI}$ from
    Eq.~(\ref{eq:nhi_final}).  Colours denote the density of points in
    this plot, {\em black dots} denote the 10th percentile of $b^2$ in
    bins of column density with bootstrap error bars. The {\em dotted
      red line} is the thermal broadening $b_T^2$ from
    Eq.~({\ref{eq:btherm}}) using Eq.~({\ref{eq:nhi_final}}) to relate
    column density to density contrast. The {\em solid black line} is
    our model for broadening of Eq.~({\ref{eq:b2_nhi}}) using the
    values $T_0$ and $\gamma$ from Fig.\ref{fig:tdelta}, $f_J=0.88$
    from Fig.~\ref{fig:b2hubble} and $f_N=0.3$ from
    Fig.~\ref{fig:nhi_rec_delta}.  The {\em black line connecting
      magenta dashes} is Eq.~(\ref{eq:b2_nhi}) using the $T$--$\Delta$
    relation measured from the simulation, rather than a power-law
    fit.  The analytic formula describes the minimum line width at a
    given column density accurately, and works even better if the
    actual $T-\Delta$ relation is used rather than a power-law fit.}
  \label{fig:broadening_nhi}
\end{figure}

Our description of line broadening so far used $b$ and $\Delta$
to relate temperature to density, yet density contrast $\Delta$ is not directly measurable. Here we investigate whether we can use the 
column density $N_{\rm HI}$ of the line instead. Following the reasoning of \cite{schaye2001}, we make the {\em Ansatz} that the column density, density, and extent of the line are related by
\begin{equation}
N_{\rm HI} \sim  n_{\rm HI} f_J \lambda_J \,.
\label{eq:nhiapprox}
\end{equation}
For highly-ionised gas in photo-ionisation equilibrium
\begin{eqnarray}
  {n_{\rm HI}\over n_{\rm H}} & = & \alpha_{\rm HII} n_e \Gamma^{-1} \\
             &\approx & 5.6\times 10^{-6}\,
             \left({n_{\rm H}\over 1.2\cdot~10^{-5}~{\rm cm^{-3}}}\right)\nonumber\\
             &\times&\left({T\over 10^4~{\rm K}}\right)^{-0.76}\,
             \left({\Gamma\over 10^{-12}~{\rm s}}\right)^{-1}\,,\nonumber
\end{eqnarray}
where 
\begin{eqnarray}  
  \alpha_{\rm HII}& = &  \alpha_0\,
  \left({T\over 10^4{\rm K}}\right)^{-0.76}\\
  \alpha_0 &= & 4\times 10^{-13}\,{\rm cm^3 s^{-1}}\,,\nonumber
\end{eqnarray}
is the value of the case-A recombination coefficient as
a function of temperature, $n_e$ is the electron number density, $n_{\rm
H}$ is the total hydrogen number density, and $\Gamma$ the photo-ionisation rate. Combining these with the expression of the Jeans length from
Eq.~(\ref{eq:lJ}) yields
\begin{eqnarray}
N_{\rm HI} &=& {3\sqrt{10}\over 64\pi}\,f_N\,f_J\,(2-Y)(1-Y) {\Omega_b^2\over \Omega_m^{1/2}}{\alpha_0\over\Gamma}{H_0^3\over ({\rm G}\,m_{\rm H})^2}\nonumber\\
&\times &
\left({k_{\rm B}\,10^4{\rm K}\over \mu\,m_{\rm H}}\right)^{1/2}
\left({T\over 10^4{\rm K}}\right)^{-0.26}(1+z)^{9/2}\Delta^{3/2}\nonumber\\
&\equiv & N_0\, \,{f_N\over 0.3}\,{f_J\over 0.88}\left({T\over 10^4{\rm K}}\right)^{-0.26}\Delta^{3/2}\,,
\label{eq:nhi_final}
\end{eqnarray}
where numerically
\begin{eqnarray}
  N_0 &=& 1.4\times 10^{13}\,{\rm cm}^{-2}\,
  \left({\Gamma\over 10^{-12}\,{\rm s}^{-1}}\right)\,
  \left({1+z\over 4}\right)^{9/2} 
 \nonumber\\
  &\times&  \left({\Omega_b\over 0.04825}\right)^{2}\,
  \left({\Omega_m\over 0.307}\right)^{-1/2}\,
  \left({h\over 0.6777}\right)^{3}\,.
\end{eqnarray}
Here, $f_N$ is the proportionality factor implicit in
Eq.~(\ref{eq:nhiapprox}) which we show below is $\sim 0.3$, and the
numerical value for $N_0$ is for very highly ionised ($\mu=0.59$)
primordial gas with the \cite{Planck13} cosmological parameters, and
$Y$ is the primordial helium abundance. For a power-law
temperature-density relation, $T=T_0\,\Delta^{\gamma-1}$, the hydrogen
column density depends only weakly on temperature, $N_{\rm HI} \propto
T_0^{-0.26}$, and it scales with density contrast as
$\Delta^{1.76-0.26\gamma}$.  We compare the results from
Eq.~(\ref{eq:nhi_final}) to our simulation in
Fig.~\ref{fig:nhi_rec_delta}, where we find the best fit of the
model to the 10th percentile distribution of the $N_{\rm HI}$ versus
$\Delta$, the justification of this choice is in
Appendix~\ref{app:nhi_rec}. The model fits very accurately the results
from the simulation over nearly four order of magnitude in column
density, as also shown in {\protect \cite{rahmati2013}} and {\protect
  \cite{tepper-garcia2012}}.

Finally, we combine Eq.~({\ref{eq:nhi_final}) with Eq.~(\ref{eq:bfinal}) to obtain a relation between the two
observables quantities, namely  the column density $N_{\rm HI}$ and
the lower-envelope of line widths at a given column density, $b$, in terms of underlying temperature-density relation of the gas, as

\begin{eqnarray}
b_T^2 &=& {2k_{\rm B}T_0\over m}\,N^{(\gamma-1)/(1.76-0.26\gamma)}\,\nonumber\\
b^2   &=& b_T^2\,\left[ 1 + 0.75\, \left({f_J\over 0.88}\right) N^{-1/(1.76-0.26\gamma)}\right]\,\nonumber\\
N     &\equiv & \left({N_{\rm HI}\over N_0}\right) 
\left({T_0\over 10^4{\rm K}}\right)^{0.26}\,\left({f_N\over 0.3}\right)^{-1}\,\left({f_J\over 0.88}\right)^{-1}\,.
\label{eq:b2_nhi}
\end{eqnarray}
This relation is compared to the $b-N_{\rm HI}$ distribution of the simulation in Fig.~\ref{fig:broadening_nhi}. Our model for line broadening does well in predicting the minimum line width as a function of column density, in particular predicting correctly that line width is not a monotonic function of column density. At low $\Delta$, line widths increase as Jeans smoothing becomes more important, whereas at high $\Delta$, line widths decrease due to radiative cooling. The underlying $T-\Delta$ relation of the simulation is only approximately a power law: using the relation from the simulation directly in Eq.~({\ref{eq:nhi_final}) improves the agreement between model and simulation further, in particular it might be possible to measure the temperature of denser gas, $\Delta\sim 10$. We illustrate the sensitivity of the $b-N_{\rm HI}$ relation to small changes in $T_0$ and $\gamma$, as well as the fitting parameters $f_J$ and $f_N$, in Appendix C.

Our model overestimates line broadening of very weak lines, $N_{\rm HI} \lsim 10^{12.5}$~cm$^{-2}$, yet it correctly predicts line widths at low density contrast (see Fig.~\ref{fig:broadening}). In Appendix~A we show that this is because line width biases how well a line of given column can be associated with a given density contrast. In practise these lines are too weak to be detected in observed spectra.

\section{Conclusions}
We have presented an analytic description of the broadening of lines in the Lyman-$\alpha$ forest, in terms of the temperature, Jeans smoothing and density of the absorbers. Identifying individual absorbers
in a spectrum as spectral stretches between two minima in the optical
depth, we calculated the line width,
$b$, from the second derivative of the optical depth at the local
maximum in $\tau$, $b^2\equiv -2\tau/\tau''$ (Eq.\ref{eq:b}). We
argued that the thermal width  of a line, $b_T$, combined with its width due to Jeans smoothing, $b_\rho$, sets a lower-limit on the total line width, $b^2\gsim b_T^2+b_\rho^2$.
We derived an expression for $b_\rho$ that depends on the temperature
$T$ and the density contrast $\Delta$ of the absorber, which smoothly
interpolates between the filtering length interpretation of
line-widths from \cite{huignedin1997} at low $\Delta$, and the near
hydrostatic equilibrium model of \cite{schaye2001} at higher
$\Delta$. This model gives a good description of the lower values of
$b$ at given $\Delta$ (Fig.~\ref{fig:broadening})~. We also discussed
the origin of the lines that are significantly wider than the lower
envelope, arguing that they result from the clustering of several individual density maxima that cannot be distinguished in the spectrum due to thermal broadening and Jeans smoothing. Such clustering is not captured by our quasi-linear model.

Based on the model of \cite{schaye2001}, we write an expression
relating column density to density contrast.
Combining this with our model for line broadening yields an analytic
expression for the lower envelope of $b$ as a function of column
density, $N_{\rm HI}$ (Eq.~\ref{eq:b2_nhi}).
This model describes the lower envelope much better than one based on thermal broadening alone (Fig.~\ref{fig:broadening_nhi})~. 

In our interpretation of $b$, line widths of weaker lines are
dominated by the filtering length, which in principle depends on the
{\em thermal history} of the IGM and not just on its instantaneous
state. If this is true, then not taking this into account may affect
the interpretation of the $b-N_{\rm HI}$ cut-off in terms of the underlying temperature-density relation of the gas.

Our analytic description of the line broadening opens the way to a new
method for determining the IGM thermal state, and in particular for measuring the
\lq longitudinal\rq\ size of the absorbers (the \lq filtering
length\rq), that is complementary to what can be measured from close QSO pairs. This is
particularly interesting, because for the first time we have a method
to estimate the sizes of the clouds from a single line of
sight.

\section{Acknowledgement}
AG thanks the Astrophysical Sector in SISSA for providing computing
facilities during the realization of this paper, Carlos Frenk and
Carlton Baugh for supporting her visit at ICC in Durham University,
Matteo~Viel for supporting her visit at Osservatorio Astronomico di
Trieste, and Uro\v{s} Seljak for appointing her as a postdoctoral
fellow at Ewha Womans University. This work has been partially done
during her permanence at the Institute for Early Universe at Ewha
Womans University, and it was supported by the WCU grant no. R32-10130
and the Research fund no. 1-2008-2935-001-2 by Ewha Womans
University. AG thanks Jeroen Franse for the English revision of part
of the manuscript.  She also thanks Fedele Lizzi for being her mentor
over many years.  JS acknowledges support from the European Research
Council under the European Union's Seventh Framework Programme
(FP7/2007-2013)/ERC Grant agreement 278594-GasAroundGalaxies.  This
work was supported by the Science and Technology Facilities Council
[grant number ST/F001166/1], and by the Interuniversity Attraction
Poles Programme initiated by the Belgian Science Policy Office ([AP
  P7/08 CHARM], and used the DiRAC Data Centric system at Durham
University, operated by the Institute for Computational Cosmology on
behalf of the STFC DiRAC HPC Facility (www.dirac.ac.uk). This
equipment was funded by BIS National E-infrastructure capital grant
ST/K00042X/1, STFC capital grant ST/H008519/1, and STFC DiRAC
Operations grant ST/K003267/1 and Durham University. DiRAC is part of
the National E-Infrastructure. The data used in the work is available
through collaboration with the authors.  \bibliographystyle{mn2e}
\bibliography{main}

\appendix

\section{Biases in the $N_{\rm HI}$--$\Delta$ relation}
\label{app:nhi_rec}
\begin{figure}
  \centering
  \includegraphics[width=\columnwidth]{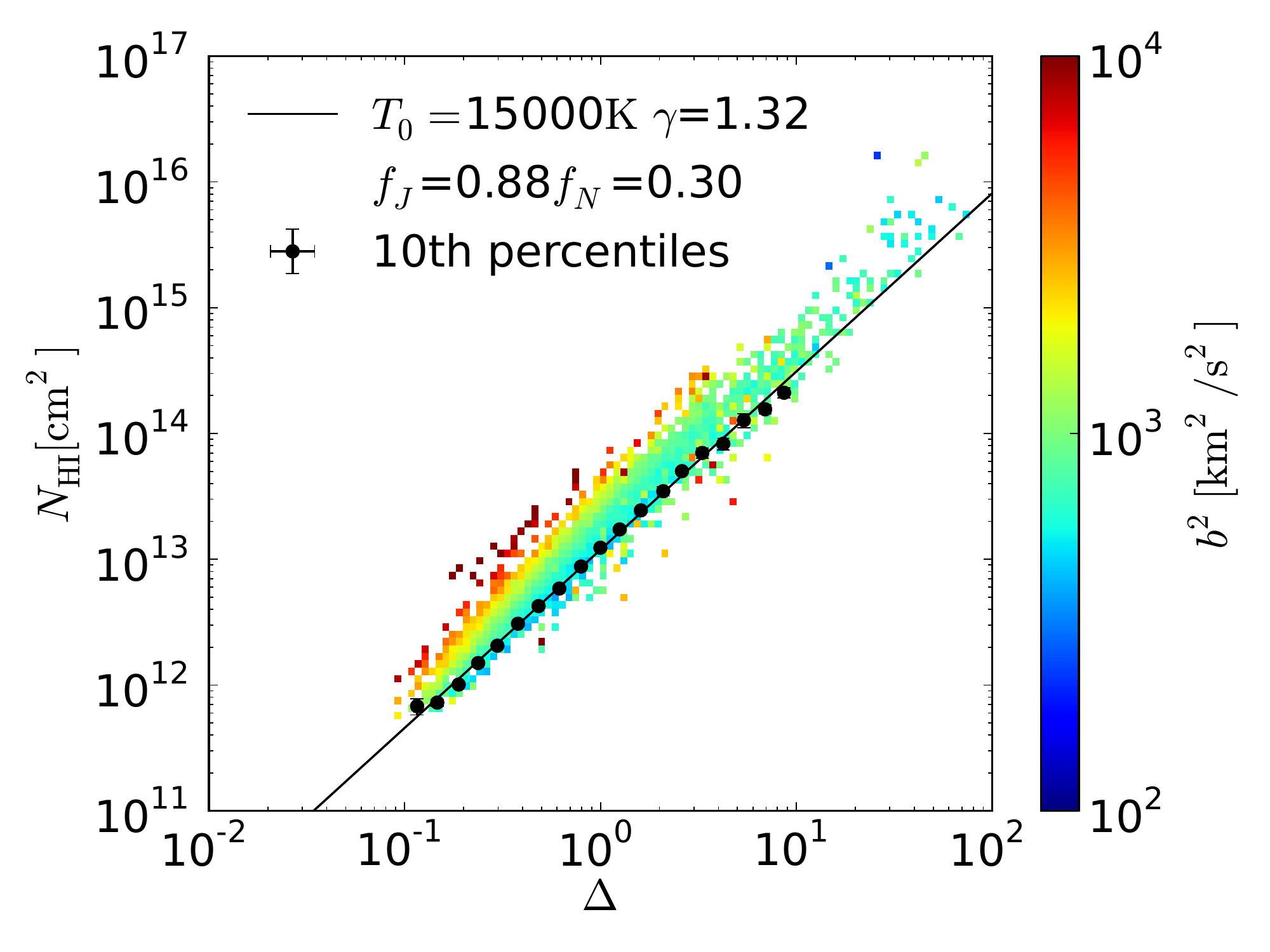}
  \caption{Density contrast - column density $\Delta$--$N_{\rm HI}$ relation for lines from the {\sc reference} model, coloured according to their line width, $b$; 
{\em black dots} denote the 10th percentile lowest value of column-density at given density contrast. 
Line width $b$ and column density $N_{\rm HI}$ are calculated from Eqs.~(\ref{eq:b}-\ref{eq:nhi}), respectively. The {\em solid black line} is the analytic relation from 
Eq.~(\ref{eq:nhi_final}). At given $\Delta$, broader lines have {\em higher} column density, illustrating that line width biases the column density associated with a given density contrast.
  }
  \label{fig:nhirec_gradient}
\end{figure}
\begin{figure}
  \centering \includegraphics[width=\columnwidth]{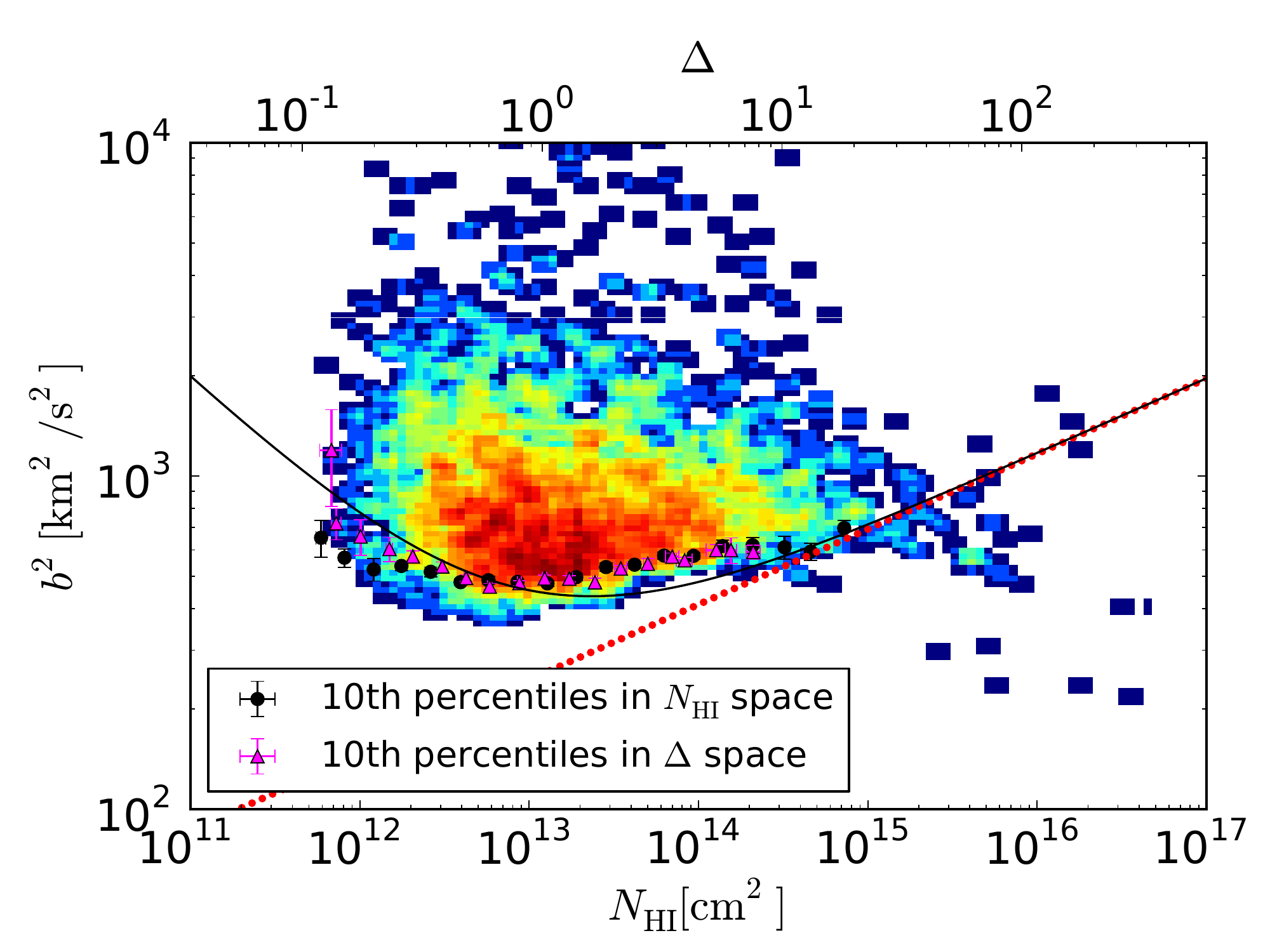}
  \caption{Column-density line width $N_{\rm HI}$-$b$ relation for the
    {\sc reference} model, colours are a measure of the density of
    points in the plot; the {\em black line} is the model illustrated
    in Fig.~\ref{fig:broadening_nhi}; the {\em dotted red} line is the
    thermal width. {\em Black dots} denote the 10th percentile lowest
    values of $b$ for bins in column density, {\em magenta triangles}
    denote the 10th percentile lowest values of $b$ for bins in
    density contrast $\Delta$, using Eq.~(\ref{eq:nhi_final}) to
    convert $\Delta$ to $N_{\rm HI}$. The bias that line width
    introduces in this relation causes the black dots and red dots to
    diverge at low $N_{\rm HI}$.  }
\label{fig:broadening_nhi_orig}
\end{figure}

\begin{figure}
  \centering
  \includegraphics[width=\columnwidth]{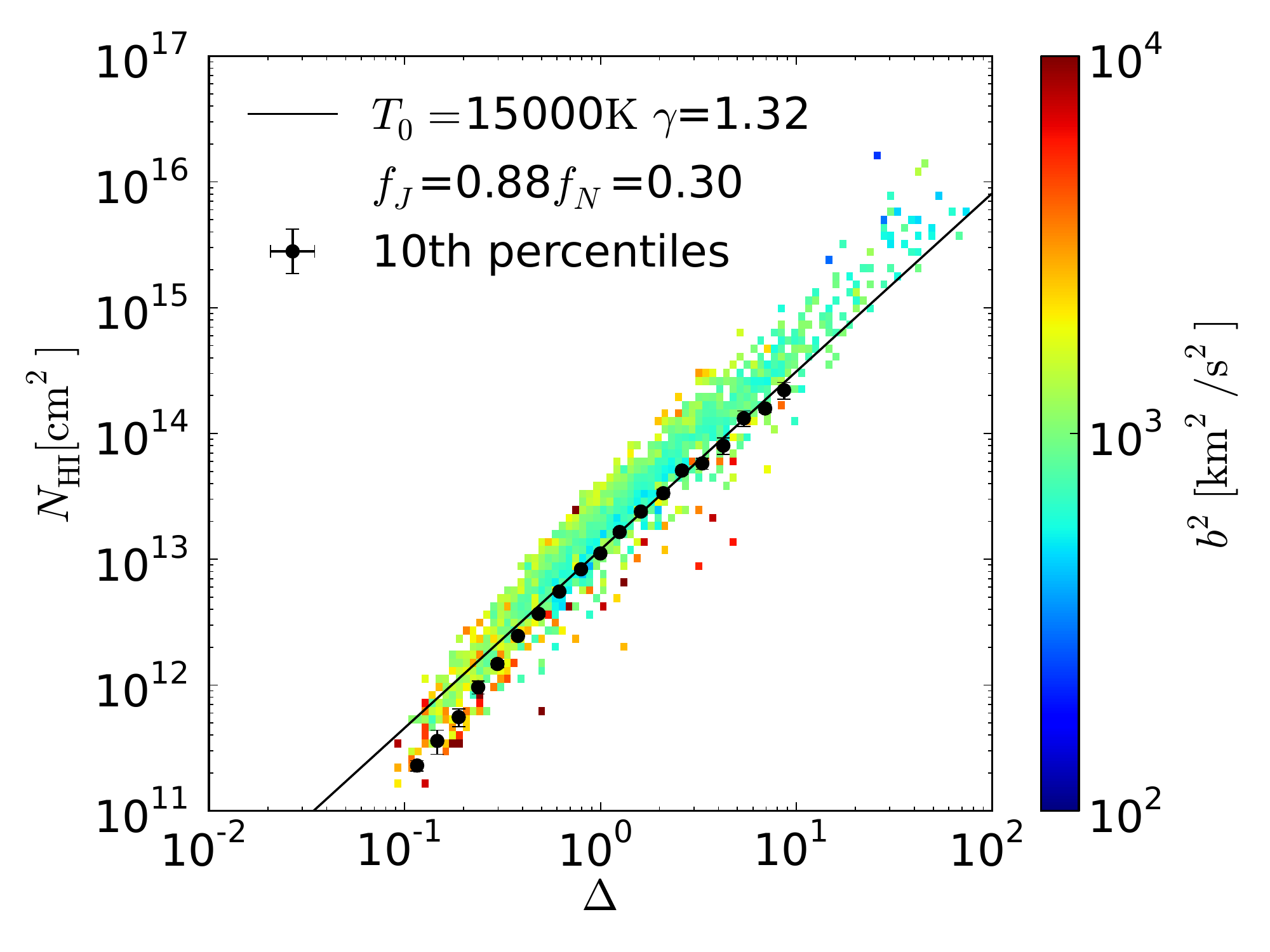}
  \caption{Same as Fig.~\ref{fig:nhirec_gradient}, but with $N_{\rm HI}$
    computed from Eq.~(\ref{eq:nhiint}). For the low-density lines the
    neutral hydrogen column density deviates from
    Eq.~(\ref{eq:nhi_final}) due to the blending of the faint
    lines.}
  \label{fig:nhi_gradient}
\end{figure}
\begin{figure}
  \centering
  \includegraphics[width=\columnwidth]{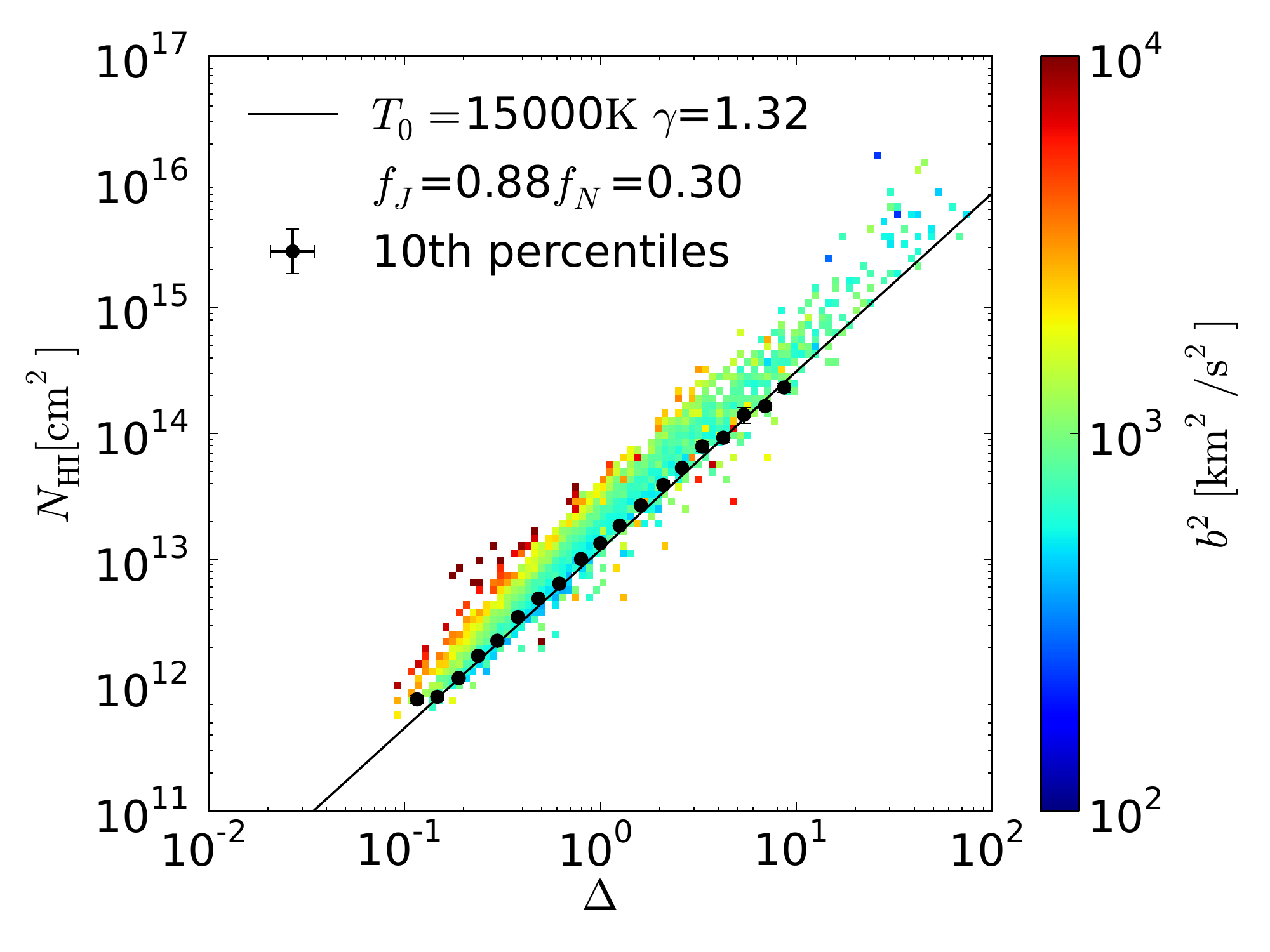}
  \caption{Same as Fig.~\ref{fig:nhirec_gradient} and \ref{fig:nhi_gradient}, but with $N_{\rm HI}$ computed from Eq.~(\ref{eq:nhiiint}). The distribution $\Delta$--$N_{\rm HI}$
    looks similar to the one shown in
    Fig.~\ref{fig:nhirec_gradient}. }
  \label{fig:nhi_corrected}
\end{figure}

In Fig.~\ref{fig:nhirec_gradient} we plot $\Delta$--$N_{\rm HI}$ for
lines, with $N_{\rm HI}$ computed from Eq.~(\ref{eq:nhi}), and
individual lines coloured with the value of the broadening $b$,
calculated from Eq.~(\ref{eq:b}). Line width biases the value of
column density, in the sense that the analytic model of
Eq.~(\ref{eq:nhi_final}) (black line in the figure) yields {\em lower}
values of $N_{\rm HI}$ than Eq.~(\ref{eq:nhi}) for broad lines. The
effect of this bias on the $N_{\rm HI}$-$b$ relation is illustrated in
Fig.~\ref{fig:broadening_nhi_orig}. Our model for line broadening
(black line) works better when lines are characterised by their
density contrast (red dots) than by column density (black dots),
especially for weak lines. This is because relation
Eq.~(\ref{eq:nhi_final}) is fitted to narrower lines.

Here we also show that Eq.~(\ref{eq:nhi}) is equivalent to computing
$N_{\rm HI}$ with the integral of the optical depth, as it is often
done in the literature.  In
Fig.~\ref{fig:nhi_gradient} we show $\Delta$--$N_{\rm HI}$, with
$N_{\rm HI}$ computed from
\begin{eqnarray}
  N_{\rm HI} & = & \frac{1}{\sigma_0 c} \int_{v_1}^{v_2} \tau \, dv \, ,
\label{eq:nhiint}
\end{eqnarray}
where $v_1$ and $v_2$ are the extremes of the line. If we compare
Fig.~\ref{fig:nhi_gradient} with Fig.~\ref{fig:nhirec_gradient}, we
can see that Eq.~(\ref{eq:nhiint}) underestimates the neutral hydrogen column density for the low-density lines.

When lines are blended, the column density computed from Eq.~(\ref{eq:nhiint}) may become smaller and smaller as a shorter velocity interval is associated with the line (i.e. when $v_1\approx v_2$). We can estimate the contribution of these wings, and compute a corrected column-density
\begin{eqnarray}
  N_{\rm HI} & = & \frac{1}{\sigma_0 c} \int_{-\infty}^{\infty} \tau \, dv \, ,
\label{eq:nhiiint}
\end{eqnarray}
by extrapolating a Gaussian fit to the line. Such an extrapolated
density is closer to what is done in Voigt-profile fitting.  Using
Eq.~(\ref{eq:nhiiint}) results in a $b-N_{\rm HI}$ distribution which
is very similar from that obtained in Fig.~\ref{fig:nhirec_gradient},
as shown in Fig.~\ref{fig:nhi_corrected}.

\section{Determination of $\lowercase{f}_J$}
\label{app:fj}
\begin{figure*}
  \centering
  \includegraphics[width=\textwidth]{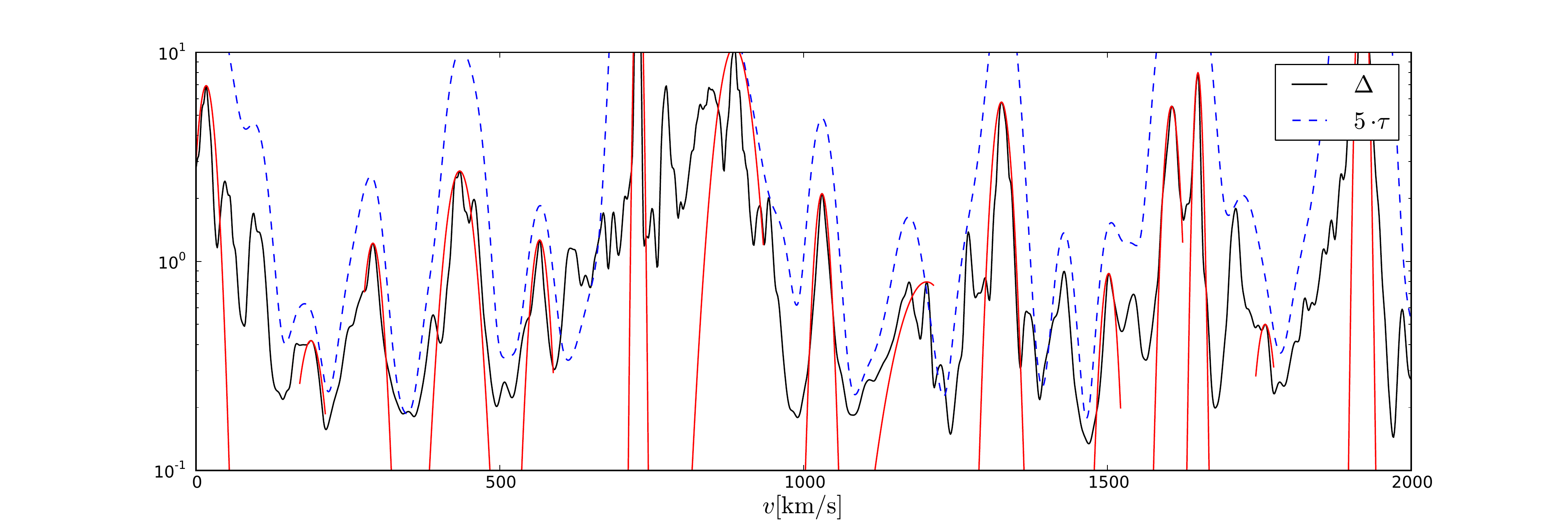}
  \caption{Density profile of absorbers in the absence of peculiar
    velocities ({\em black line}) for a mock spectrum calculated from
    the {\sc reference} model; {\em dashed blue line} is the scaled
    optical depth for comparison. {\em Red lines} are Gaussian
    profiles from Eq.~(\ref{eq:Gr}), where the width is such that the
    integral of the density profile of the Gaussian is the same as the
    one of the multi-peaked absorber. The widths of the red Gaussians
    are often a good approximation to the extent of an absorber.  }
  \label{fig:rsfitting}
\end{figure*}

In this Appendix we illustrate that our estimate for the extent of absorbers in real space
indeed is a good approximation for their actual size. We begin by neglecting peculiar velocities (i.e. generate mock spectra {\em after} zeroing all peculiar velocities), so that we can unambiguously identify the real-space density structure that gives rise to a given line; we demonstrated in the main text that peculiar velocities have little impact on line-widths.  Often individual absorption lines correspond to more than one density peak. Such clustering of peaks sometimes gives rise to very wide lines, as we argued in the text.

In Fig.~\ref{fig:rsfitting} we show how we have computed the
broadening $b_\Delta$ in Fig.~\ref{fig:b2hubble} from the density
profile of the multi-peaked absorber.  The red lines are Gaussian
profiles as in Eq.~(\ref{eq:Gr}), with $v_0$ and $\rho_0$ respectively
the velocity and the density of the local maximum in optical
depth. The width $b_\Delta$ is such that the integral of the Gaussian
density profile is the same as the integral of the density profile of
the absorbers. Even though there is often no unique way of associating
a \lq size\rq\ with such complex absorbers, we believe $b_\Delta$ is
often a good approximation of the extent over which the density
contrast of an absorber is significant.

\section{Degeneracy of parameters in the line broadening relation}
For completeness we illustrate the sensitivity of Eq.~(\ref{eq:bfinal}) to changes in
the $T-\Delta$ relation (varying $T_0$ and $\gamma$; Fig.\ref{fig:broadening_nhi_f1}), and varying the fitting parameters $f_J$ and $f_N$ (Fig.~\ref{fig:broadening_nhi_f2})

\begin{figure}
  \centering \includegraphics[width=\columnwidth]{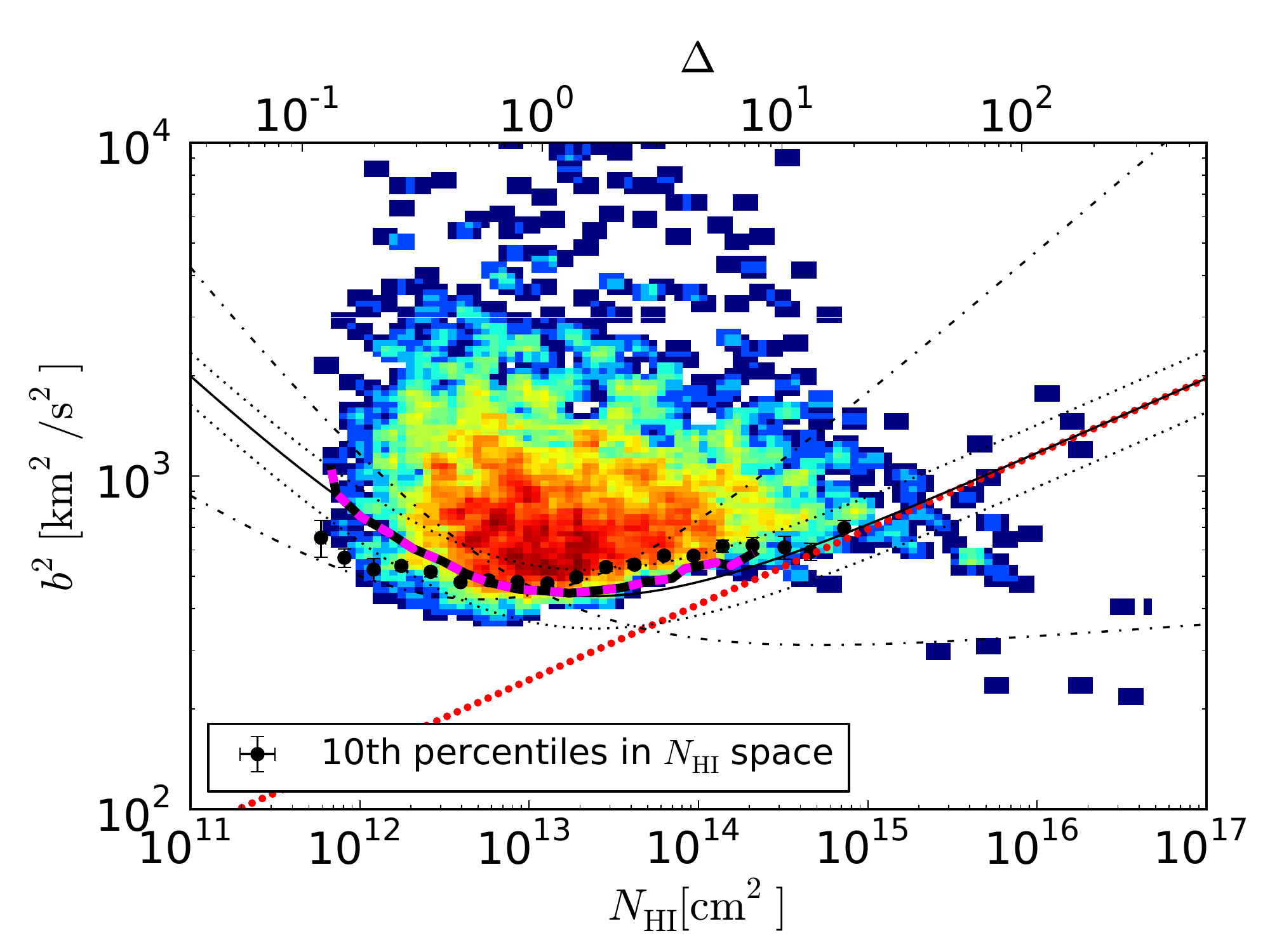}
  \caption{Same as figure~\ref{fig:broadening_nhi}.  The dashed
    (dashed-dot) lines are the model from Eq.~\ref{eq:b2_nhi} with
    $T_0$ ($\gamma$) changed by $\pm 20$\%. }
  \label{fig:broadening_nhi_f1}
\end{figure}

\begin{figure}
  \centering \includegraphics[width=\columnwidth]{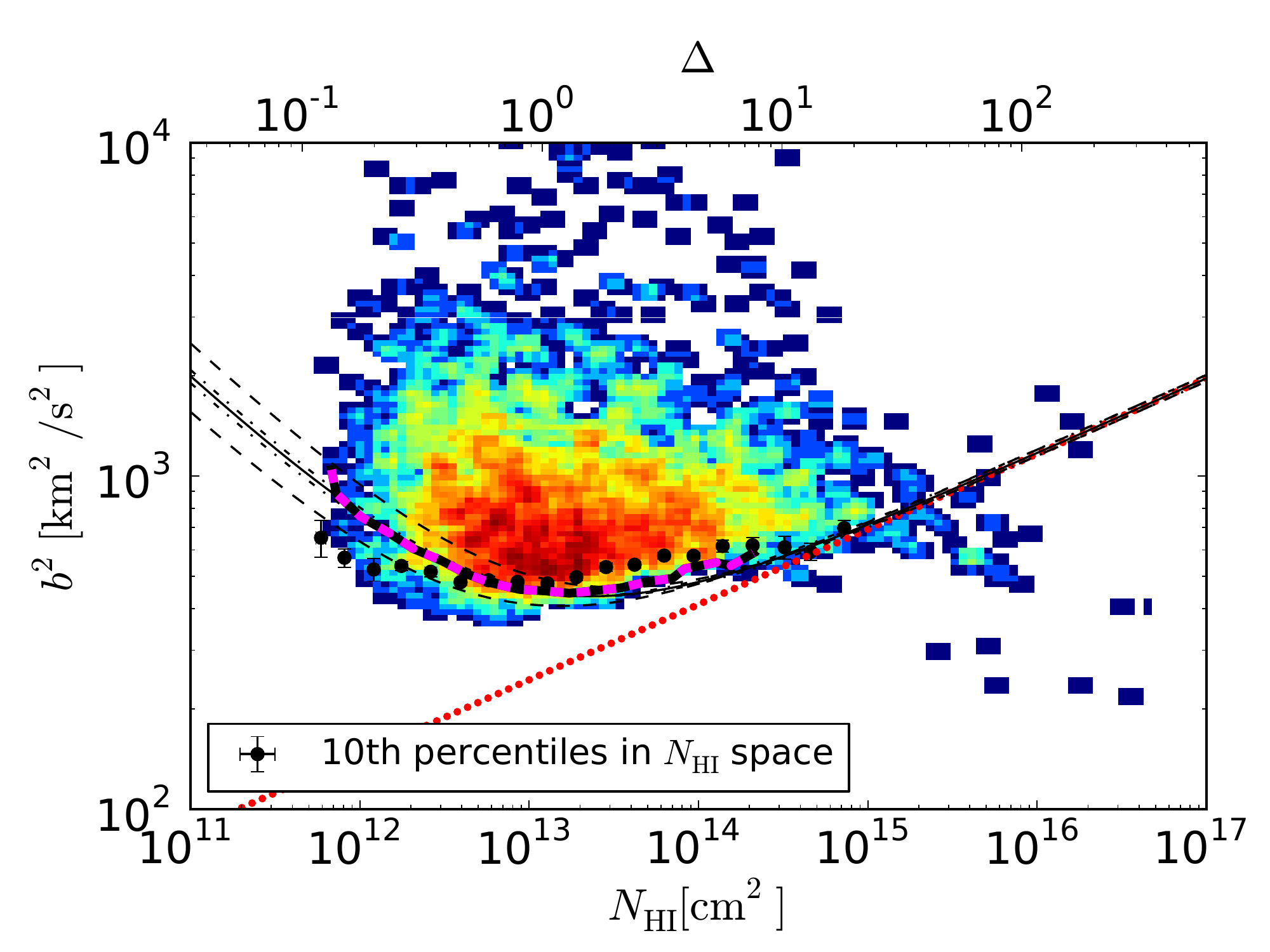}
  \caption{Same as figure~\ref{fig:broadening_nhi}. The dashed
    (dashed-dot) lines are the model from Eq.~\ref{eq:b2_nhi} with
    $f_J$ ($f_N$) changed by $\pm 10$\%. }
  \label{fig:broadening_nhi_f2}
\end{figure}

\end{document}